\documentclass[12pt]{article}

\usepackage{epsfig}
\usepackage{graphicx}
\usepackage{amssymb,amsmath}
\usepackage{hyperref}
\usepackage[nosort]{cite}
\hypersetup{
    colorlinks=true,
    linkcolor=blue,
    citecolor=red,
}

\begin{document}

\begin{center}
\Large{\bf Scalar field inflation driven by a modification of the Heisenberg algebra} \vspace{0.5cm}

\large  H. Garc\'{\i}a-Compe\'an\footnote{e-mail address: {\tt
compean@fis.cinvestav.mx}}, D. Mata-Pacheco\footnote{e-mail
address: {\tt dmata@fis.cinvestav.mx}}

\vspace{0.3cm}

{\small \em Departamento de F\'{\i}sica, Centro de
Investigaci\'on y de Estudios Avanzados del IPN}\\
{\small\em P.O. Box 14-740, CP. 07000, Ciudad de M\'exico, M\'exico}\\

\vspace*{1.5cm}
\end{center}

\begin{abstract}
We study the modifications induced on scalar field inflation produced by considering a general modification of the Heisenberg algebra. We proceed by modifying the Poisson brackets on the classical theory whenever the corresponding quantum commutator is modified. We do not restrict ourselves to a specific form for such modification, instead we constrain the functions involved by the cosmological behaviour of interest. We present whenever possible the way in which inflation can be realized approximately via three slow roll Hubble parameters that depend on the standard slow roll parameters in a very different form than in the usual case and that can be less restrictive. Furthermore we find a general analytical solution describing an expanding universe with constant Hubble parameter that generalizes the standard cosmological constant case by restricting the form of the modification of the Heisenberg algebra. It is found that even if such modification can be neglected in some limit and the cosmological constant is set to zero in that limit, the exponential expansion is present when the modification is important. Thus an appropriate modification of the Heisenberg algebra is sufficient to produce an exponentially expanding universe without the need of any  other source.

\vskip 1truecm

\end{abstract}

\bigskip

\newpage

\section{Introduction}
\label{S-Intro}

The Heisenberg Uncertainty Principle (HUP) was one of the first features of quantum theories that changed the paradigm when first encountered with quantum behaviour. However as the quantization procedure were pursued in a wider types of systems a generalization of this principle was found to be necessary motivated within different scenarios. At the present time it is believed that a quantum theory of gravity, whatever its form must contain a minimum measurable length \cite{STLength,STLength2,Maggiore:1993rv,Scardigli:1999jh,Garay:1994en} which is incompatible with the HUP. Thus instead of working with a proposal of a quantum theory for gravity we can modify the uncertainty principle by modifying the Heisenberg algebra in the quantum theories that we already have to incorporate this behaviour and thus gain some insight to the modifications that we expect are going to arrive with the full quantum gravity theory without having to deal with theories that are far more complicated. This modification has been called a Generalized Uncertainty Principle (GUP). The easiest way to incorporate such minimum length is by considering that the GUP takes the form\footnote{Throughout this work we will use units where $\hbar=c=1$.} \cite{Kempf:1994su}
	\begin{equation}\label{SimplestGUP}
		[x,p]=i(1+\gamma^2p^2) ,
	\end{equation}
where $\gamma$ is a small parameter which will play the role of the minimal measurable length. However in general it can be shown that by proposing a GUP in the form \cite{Shababi:2019ysa}
	\begin{equation}
		[x,p]=iF(p^2) ,
	\end{equation}
we can have a minimal measurable length if  the function $F(x^2)/x$ has a minimum once we choose states such that $\langle p\rangle=0$. Thus a wide variety of GUPs have been proposed in the literature, generalizing (\ref{SimplestGUP}) in terms of geometric series  \cite{Pedram:2011gw}
	\begin{equation}\label{GUPSeries}
		[x,p]=\frac{i}{1-\gamma^2p^2}.
	\end{equation}
This form is also compatible with loop quantum cosmology and the theory of Doubly Special Relativity. Functions that coincides with (\ref{SimplestGUP}) on a Taylor series at first order have also been proposed such as one inspired by non-commutative Snyder space \cite{Snyder:1946qz,Battisti:2008du}
	\begin{equation}\label{GUPSR}
		[x,p]=i\sqrt{1-\gamma p^2} ,
	\end{equation}
or even an exponential proposal in the form \cite{Miao:2013wua}
	\begin{equation}
		[x,p]=ie^{\gamma p^2} .
	\end{equation}

However, the look for a minimal measurable length is not the only motivation for a generalization of the  HUP. It has been found that when the quantization procedure takes place in a dS or AdS space the uncertainty principle has to be modified in order to incorporate a maximum distance given by the dS (or AdS) radius, this has been called the Extended Uncertainty Principle (EUP) \cite{Park:2007az,Mignemi:2009ji,Gine:2020izd}. Modifying the uncertainty principle by incorporating a dependence on the momentum as well as the position have also been performed called Generalized Extended Uncertainty Principle (GEUP) in \cite{Park:2007az,Mignemi:2009ji}. On the other hand in the context of the Doubly Special Relativity theory it has also been found that a modification of the HUP is needed \cite{Magueijo:2001cr,Cortes:2004qn} which leads to a form that is compatible with string theory and black hole physics as well \cite{Magueijo:2004vv,Ali:2009zq}. Furthermore considering noncommutativity on the variables of spacetime can also lead to a modification of the HUP which are of great interest as well (for some reviews, see
\cite{Douglas:2001ba,Szabo:2001kg}). Therefore, there are many scenarios that motivates the generalization of the HUP and can take many forms. In the present article however we will view any generalization of the uncertainty principle in the form of a modification of the Heisenberg algebra as a GUP for a simpler notation. 

A cosmological setup represents a natural interesting scenario to study the implications of these generalizations since it is expected that the early universe provides a natural framework to study quantum gravity phenomenology. The implication of considering a GUP in the variables of superspace when studying the Wheeler-De Witt equation have been performed before, for example see \cite{Garcia-Compean:2001jxk,Vakili:2007yz,Vakili:2008tt,Kober:2011uj,Zeynali:2012tw,Faizal:2014rha,Faizal:2015kqa,Garattini:2015aca,Gusson:2020pgh,GUP,GUPHL}. On the other hand the modifications arriving from considering a GUP on the scalar field modes in the CMBR spectrum have been studied previously on \cite{Kempf:2000ac,Ashoorioon:2004vm,Ashoorioon:2004wd,Ashoorioon:2005ep}, where it is shown that there appears an ambiguity regarding the action for the scalar modes. Furthermore modifications to the Friedmann equation can be obtained using a thermodynamical approach by considering the implications of the GUP, this has been carried out in \cite{Cai:2005ra,Zhu:2008cg,Giardino:2020myz} where it was found that for the forms of the GUPs considered they can not distinguish the behaviour with the standard HUP case. Moreover the role of an scalar field on cosmology can be studied by considering the effects of the GUP in the action of the scalar field as in \cite{Paliathanasis:2015cza,Giacomini:2020zmv}, leading to results consisting of an exponentially expanding universe. A noncommutative approach have also been proposed in \cite{Bina:2007wj,Leon:2022oeo,Perez-Payan:2011cvf,Toghrai:2021nuz}.  A similar treatment exploring the classical as well as the quantum implications of a GUP and polymer quantum mechanics on cosmology have been performed in \cite{Barca:2019ane,Barca:2021epy,Barca:2023epu}. However we can also pursue modifications to the Friedmann equation for the GUPs that we are interested in implementing a classical limit in which the modification of the Heisenberg algebra of commutators implies a modification of the Poisson brackets of the classical system. In this way by considering a universe filled with a perfect fluid in \cite{Battisti:2008du,Ali:2014hma,Moumni:2020uki,Atazadeh:2016yeh} the modified Friedmann equations were investigated. It was found that the resulting cosmology is compatible with a cosmological bounce or even with an static universe depending on the form of the GUP. As far as we know the generalization of this treatment by taking into account an scalar field with an arbitrary potential has not been worked out previously in the literature. This is the main goal of the present work.

The article is organized as follows. In Section \ref{S-NoScalarField} we will review the way in which the modified Friedmann equations by the GUP are obtained when the only source of the gravitational field is a perfect fluid. We then generalize this treatment by considering explicitly an scalar field with an arbitrary potential in Section \ref{S-ScalarField}. Since we do not want to restrict ourselves to specific forms for the GUP we propose general forms that the GUP can take and study each generalization in Sections \ref{S-Case1}-\ref{S-Case4}. Finally we present our Final Remarks in Section \ref{S-FinalRemarks}.

\section{Perfect fluid cosmology with a GUP}
\label{S-NoScalarField}
Let us begin with a review of how the modification from the Heisenberg algebra derived from the GUP has been studied in a classical limit in order to modify the Friedmann equations at the classical level when the only source of the gravitational field is described by a perfect isotropic fluid with energy density $\rho$ and pressure $p$. The gravitational part of the action will be described by the Einstein-Hilbert action and we will consider the FLRW flat metric, which is commonly used in inflationary cosmological scenarios, described by
	\begin{equation}\label{FLRWMetric}
		ds^2=-N^2(t)dt^2+a(t)^2\left(dx^2+dy^2+dz^2\right) ,
	\end{equation}
where $N(t)$ is the lapse function and $a(t)$ is the scale factor. In the following we will always choose $N=1$. The hamiltonian constraint that follows from the standard formulation is written as
	\begin{equation}\label{SistNSF1}
		\mathcal{H}=-\frac{\kappa}{12a}P^2_{a}+a^3\rho\simeq0 ,
	\end{equation}
where $\kappa=8\pi G$ and $P_{a}$ is the canonical momentum associated to $a$. For the total hamiltonian we must add the momentum constraints but in the cases that we will consider these terms will be unimportant and thus we can ignore them and use the hamiltonian constraint as the full hamiltonian. Following the standard formulation the equations of motion will be given by
	\begin{equation}\label{SistNSF2}
		\dot{a}=\{a,\mathcal{H}\}=\frac{\partial \mathcal{H}}{\partial P_{a}} , \hspace{1cm} \dot{P_{a}}=\{P_{a},\mathcal{H}\}=-\frac{\partial\mathcal{H}}{\partial a},
	\end{equation}
followed by the standard conservation law
	\begin{equation}\label{SistNSF3}
		\rho+3H(\rho+p)=0 ,
	\end{equation}
where $H=\frac{\dot{a}}{a}$ is the Hubble parameter. The first equation in (\ref{SistNSF2}) is just the definition of the momentum $P_{a}$. When substituting this value in the hamiltonian constraint (\ref{SistNSF1}) we obtain the standard Friedmann equation
	\begin{equation}\label{FriedmannStandard}
		H^2=\frac{\kappa}{3}\rho ,
	\end{equation}
The second equation in (\ref{SistNSF2}) leads to the second Friedmann equation after using (\ref{SistNSF3}) and can be derived from (\ref{FriedmannStandard}).

As it is well known there is a relation between the classical Poisson brackets of any two variables $\{\cdot,\cdot\}$ and the commutator of their corresponding quantum operators $[\cdot,\cdot]$. This relation is stated as
	\begin{equation}\label{PoissionConm}
		\{\cdot,\cdot\}\leftrightarrow i[\cdot,\cdot] .
	\end{equation}
In this case where we only have one coordinate in the Wheeler's superspace, that is, the hamiltonian is described only by the general coordinate $a(t)$, we can consider a modification to the commutator of $a$ with its corresponding momentum and using the relation (\ref{PoissionConm}) we will obtain modifications of the classical equations of motion and thus the Friedmann equations will be modified. Let us consider a GUP that can lead to a minimum value for the position, that is let us consider\footnote{In this work we will use the same notation for the classical variables as well as for the corresponding quantum operators.}
	\begin{equation}
		[a,P_{a}]=iF(P^2_{a}) ,
	\end{equation}
where $F$ is any function that must depend on a parameter $\gamma$ such that $\lim_{\gamma\to0}F=1$. Therefore, we will modify the standard Poisson brackets as
	\begin{equation}\label{NSFConmu}
		\{a,P_{a}\}\rightarrow\{a,P_{a}\}F .
	\end{equation}

In general we can express the bracket of any two functions $A$ and $B$ for a hamiltonian system with coordinates $x_{i}$ and momenta $p_{i}$ as
	\begin{equation}\label{PoissonGeneral}
		\{A,B\}=\left(\frac{\partial A}{\partial x_{i}}\frac{\partial B}{\partial p_{j}}-\frac{\partial A}{\partial p_{i}}\frac{\partial B}{\partial x_{j}}\right)\{x_{i},p_{j}\} .
	\end{equation}
Therefore, in this case we obtain from (\ref{NSFConmu}) and (\ref{PoissonGeneral}) that the equations of motion are given by
	\begin{equation}\label{SistNSFGUP1}
		\dot{a}=\frac{\partial H}{\partial P_{a}}F(P^2_{a}) ,\hspace{1cm}
		\dot{P}_{a}=-\frac{\partial H}{\partial a}F(P^2_{a}) .
	\end{equation}
The form of the hamiltonian constraint (\ref{SistNSF1}) is not modified. However we note from the first equation in (\ref{SistNSFGUP1}) that the definition of the momentum $P_{a}$ is modified and thus even if the hamiltonian constraint has the same form, the resulting Friedmann equation will be different for every function $F$. This expression represents an equation that must be solved for the momentum and then substituting it back in the hamiltonian constraint the Friedmann equation can be obtained, thus this procedure can not be carried out in general, we must know the specific form of the $F$ function to proceed. Let us consider some functions that have been used previously in the literature:
\begin{itemize}
	\item Let us consider first the GUP in the form of (\ref{GUPSR}) then the Poisson brackets will be
	\begin{equation}
		\{a,P_{a}\}=\sqrt{1-\gamma P^2_{a}} ,
	\end{equation}
	with this the first equation in (\ref{SistNSFGUP1}) is
	\begin{equation}\label{GUP1EOM}
		\dot{a}=-\frac{\kappa N}{6a}P_{a}\sqrt{1-\gamma P^2_{a}} ,
	\end{equation}
	which leads into a quartic equation for $P_{a}$ that can be solved and then substituted back into (\ref{SistNSF1}) to obtain the modified Friedmann equation \cite{Battisti:2008du} 
	\begin{equation}\label{ModifiedFriedmann1}
		\left(\frac{\dot{a}}{a}\right)^2=\frac{\rho \kappa}{3}\left[1-\frac{12\gamma\rho}{\kappa}a^4\right] .
	\end{equation}
	\item In view of the  GUP compatible with DSR for a single coordinate we obtain
	\begin{equation}
		\{a,P_{a}\}=1-2\gamma P_{a} ,
	\end{equation}
	the redefinition of the momentum leads to a quadratic equation that can be solved to get the modified Friedman equation \cite{Ali:2014hma}
	\begin{equation}\label{ModifiedFriedmann2}
		\left(\frac{\dot{a}}{a}\right)^2=\frac{\kappa\rho}{3}\left[1-8\gamma\sqrt{\frac{3\rho}{\kappa}}a^2+\frac{48\gamma^2\rho}{\kappa}a^4\right] .
	\end{equation}
	\item Bearing in mind now the GUP of the form of (\ref{GUPSeries}) we have
	\begin{equation}
		\{a,P_{a}\}=\frac{1}{1-\gamma P^{2}_{a}} .
	\end{equation}
	we then obtain a quadratic equation for $P_{a}$ that leads to the modified Friedmann equation \cite{Moumni:2020uki}
	\begin{equation}\label{ModifiedFriedmann3}
		\left(\frac{\dot{a}}{a}\right)^2=\frac{\kappa\rho}{3}\left[\frac{12\rho\gamma}{\kappa}a^4-1\right]^{-2} .
	\end{equation}
\end{itemize}
As we can see the form of the $F$ function determines the modified Friedmann equation that we can obtain. One important aspect to remark is that all the equations obtained described a cosmological bounce, that is there is a non-zero value of $a$ for which $\dot{a}=0$ and thus they are not related to the standard inflationary paradigm. However, so far we have only considered a perfect fluid as the gravitational source, things can change significantly if we consider an scalar field with a potential. The modification arriving from considering a noncommutative approach with an scalar potential have been studied in \cite{Bina:2007wj,Toghrai:2021nuz}. On the other hand in \cite{Battisti:2008du} it is considered an scalar field with a GUP but without a potential and the analysis is simplified by a proper choice for the lapse function. Furthermore in \cite{Paliathanasis:2015cza,Giacomini:2020zmv} a numerical analysis has been performed that points towards an agreement between the GUP scenario with an scalar field and standard inflationary behaviour.  Thus in the following section we will generalize the analysis presented in this section to take into account an scalar field with a general potential.

\section{Scalar field cosmology with a GUP}
\label{S-ScalarField}
Let us consider the standard scenario firstly, that is we will consider the action for the gravitational part to be the Einstein-Hilbert action and a minimally coupled homogeneous scalar field $\phi(t)$ with its potential $V(\phi)$, that is the action is given by
	\begin{equation}
		S=\int d^4x\sqrt{-g}\left[\frac{R}{2\kappa}+\frac{\dot{\phi}^2}{2}-V(\phi)\right] .
	\end{equation}
Considering the FLRW flat metric (\ref{FLRWMetric}) the hamiltonian constraint in this case takes the form
	\begin{equation}\label{SistStandardHC}
		\mathcal{H}=-\frac{\kappa}{12a}P^2_{a}+\frac{1}{2a^3}P^2_{\phi}+a^3V(\phi)\simeq0 ,
	\end{equation}
and thus we have four equations of motion given by
	\begin{equation}\label{SistStandard1}
		\dot{a}=\{a,\mathcal{H}\}=\frac{\partial \mathcal{H}}{\partial P_{a}} , \hspace{1cm} \dot{\phi}=\{\phi,\mathcal{H}\}=\frac{\partial \mathcal{H}}{\partial P_{\phi}} ,
	\end{equation}
	\begin{equation}\label{SistStandard2}
		\dot{P_{a}}=\{P_{a},\mathcal{H}\}=-\frac{\partial \mathcal{H}}{\partial a} , \hspace{1cm} \dot{P_{\phi}}=\{P_{\phi},\mathcal{H}\}=-\frac{\partial \mathcal{H}}{\partial \phi} .
	\end{equation}
The two equations in (\ref{SistStandard1}) are just the definitions of the momenta as usual. We can obtain the Friedmann equation by substituting these values into the hamiltonian constraint (\ref{SistStandardHC}) which leads to 
	\begin{equation}\label{FriedmannStandardC}
		H^2=\frac{\kappa}{3}\left(\frac{\dot{\phi}^2}{2}+V\right) .
	\end{equation}
The second equation of (\ref{SistStandard2}) leads to the field equation of motion
	\begin{equation}\label{FieldEoMStandard}
		\ddot{\phi}+3H\dot{\phi}+V'=0 ,
	\end{equation}
where prime denotes derivative with respect the scalar field. Finally the first equation of (\ref{SistStandard2}) leads to the second Friedman equation that can be derived from (\ref{FriedmannStandardC}) and (\ref{FieldEoMStandard}).

In order to modify the standard picture just presented by considering a generalization of the Heisenberg algebra we need to proceed as before, by modifying the  Poisson brackets. However as explained in the Introduction section the modifications of the Heisenberg algebra have been motivated with many different forms looking for different behaviours. Thus our goal in this article is to present an analysis as general as possible and therefore we will not restrict ourselves a priori to a particular form for the GUP. Instead we will consider a general form of the GUP with arbitrary functions that can depend on the coordinates and the momenta and we will restrict its form only by the cosmological behaviour of interest. However, as we have noted before, the dependence on the function is important since it will change the equations that  defines the momenta, thus we have to work out different forms of generalizations separately. We will consider the following scenarios;

\begin{itemize}
	\item \textbf{Case I}. The simplest modification of the Heisenberg algebra is to consider the diagonal case
		\begin{equation}\label{DefCase1}
			[a,P_{a}]=[\phi, P_{\phi}]=iF(a,\phi) ,
		\end{equation}
	where $F(a,\phi)$ is a function that can depend only on the coordinates $a$ and $\phi$.
	\item \textbf{Case II}. If we want to consider a dependence on the momenta, the simplest way to do it is by proposing two functions, one for each coordinate, thus the second case will be defined by
		\begin{equation}\label{DefCase2}
			[a,P_{a}]=iF_{1}(a,P_{a}) , \hspace{1cm} [\phi,P_{\phi}]=iF_{2}(\phi,P_{\phi}) .
		\end{equation}
	\item \textbf{Case III}. Considering a more complicated scenario we can move on to a function that can depend on the coordinates as well as the momenta but still has a diagonal form 
		\begin{equation}\label{DefCase3}
			[a,P_{a}]=[\phi, P_{\phi}]=iF(a,\phi,P_{a},P_{\phi}) .
		\end{equation}
	\item \textbf{Case IV}. Finally, the most general scenario is when there is not a diagonal case, that is when the commutator of a coordinate with the momentum of another coordinate is also modified, that is we will consider 
		\begin{equation}\label{DefCase4}
			[a,P_{a}]=iF_{1} , \hspace{1cm} [\phi,P_{\phi}]=iF_{2} , \hspace{1cm} [a,P_{\phi}]=iG_{1} , \hspace{1cm} [\phi,P_{a}]=iG_{2},
		\end{equation}
	where in this case all functions $F_{1,2}$ and $G_{1,2}$ can depend on the momenta as well as both coordinates.
\end{itemize}
In the following subsections we will consider what are the changes implied by all the cases in the equations of motion and what implications do they have regarding the inflationary scenario.

\subsection{Case I}
\label{S-Case1}
Let us start by considering a diagonal form with no dependence on the momenta as defined in (\ref{DefCase1}). In this case the equations of motion are modified as follows
	\begin{equation}\label{SistCase1-1}
		\dot{a}=\frac{\partial \mathcal{H}}{\partial P_{a}}F=-\frac{\kappa P_{a}}{6a}F , \hspace{1cm} \dot{\phi}=\frac{\partial\mathcal{H}}{\partial P_{a}}F=\frac{P_{\phi}}{a^3}F ,
	\end{equation}
	\begin{equation}\label{SistCase1-2}
		\dot{P}_{a}=-\frac{\partial\mathcal{H}}{\partial a}F , \hspace{1cm} \dot{P}_{\phi}=-\frac{\partial\mathcal{H}}{\partial\phi}F .
	\end{equation}
We note that this is the simplest case because the redefinition of the momenta from (\ref{SistCase1-1}) is trivial, they can be expressed in terms of the coordinates in a general form as
	\begin{equation}\label{Case1RedMomenta}
		P_{a}=-\frac{6a\dot{a}}{\kappa F} , \hspace{1cm} P_{\phi}=\frac{a^3\dot{\phi}}{F} .
	\end{equation}
then, substituting back into the hamiltonian constraint (\ref{SistStandardHC})  we obtain the modified Friedmann equation
	\begin{equation}\label{Case1Friedmann}
		H^2=\frac{\kappa}{3}\left[\frac{\dot{\phi}^2}{2}+F^2(a,\phi)V\right].
	\end{equation}
From the second equation in (\ref{SistCase1-2}) we can obtain the modified equation of motion for the scalar field
	\begin{equation}\label{Case1ScalarEq}
		\ddot{\phi}+(3H-H_{F})\dot{\phi}+F^2V'=0 ,
	\end{equation}
where $H_{F}=\frac{\dot{F}}{F}$. With these modifications we can investigate how the standard picture of inflation is modified. As it is well known scalar field inflation is realized after making two approximations, in the Friedmann equation (\ref{FriedmannStandardC}) it is assumed that $\frac{\dot{\phi}^2}{2}\ll V$ and in the scalar field equation (\ref{FieldEoMStandard}) that $\ddot{\phi}\ll H\dot{\phi}$.  These approximations are justified when the two slow roll Hubble parameters are small, they are defined by
	\begin{equation}
		\epsilon_{H}=-\frac{\dot{H}}{H^2} , \hspace{1cm} \eta_{H}=-\frac{\ddot{\phi}}{H\dot{\phi}} .
	\end{equation}
In the standard scenario we can write 
	\begin{equation}\label{RelationParameters} 
		\epsilon_{H}\thickapprox\frac{3\dot{\phi^2}}{2V}\thickapprox\epsilon_{V} , \hspace{1cm} \eta_{H}\thickapprox\eta_{V}-\epsilon_{V} ,
	\end{equation}
where the symbol $\thickapprox$ denotes an approximation, $\epsilon_{V}$ and $\eta_{V}$ are the slow roll parameters defined by
	\begin{equation}\label{DefSlowRollP}
		\epsilon_{V}=\frac{1}{2\kappa}\left(\frac{V'}{V}\right)^2 , \hspace{1cm} \eta_{V}=\frac{1}{\kappa}\frac{V''}{V} ,
	\end{equation}
and they fulfil $\epsilon_{V},\eta_{V}\ll1$. Therefore by imposing these parameters to be small we immediately obtain $\epsilon_{H},\eta_{H}\ll1$ and it is ensured that $V\thickapprox$ constant and that $H\thickapprox\frac{\kappa V}{3}$. Thus we have exponential expansion and the approximation is justified. Then in the standard case we can justify the approximation and obtain an exponential expansion by restricting the potential by the slow roll parameters. 

\begin{figure}[h]
		\centering
		\includegraphics[width=0.8\textwidth]{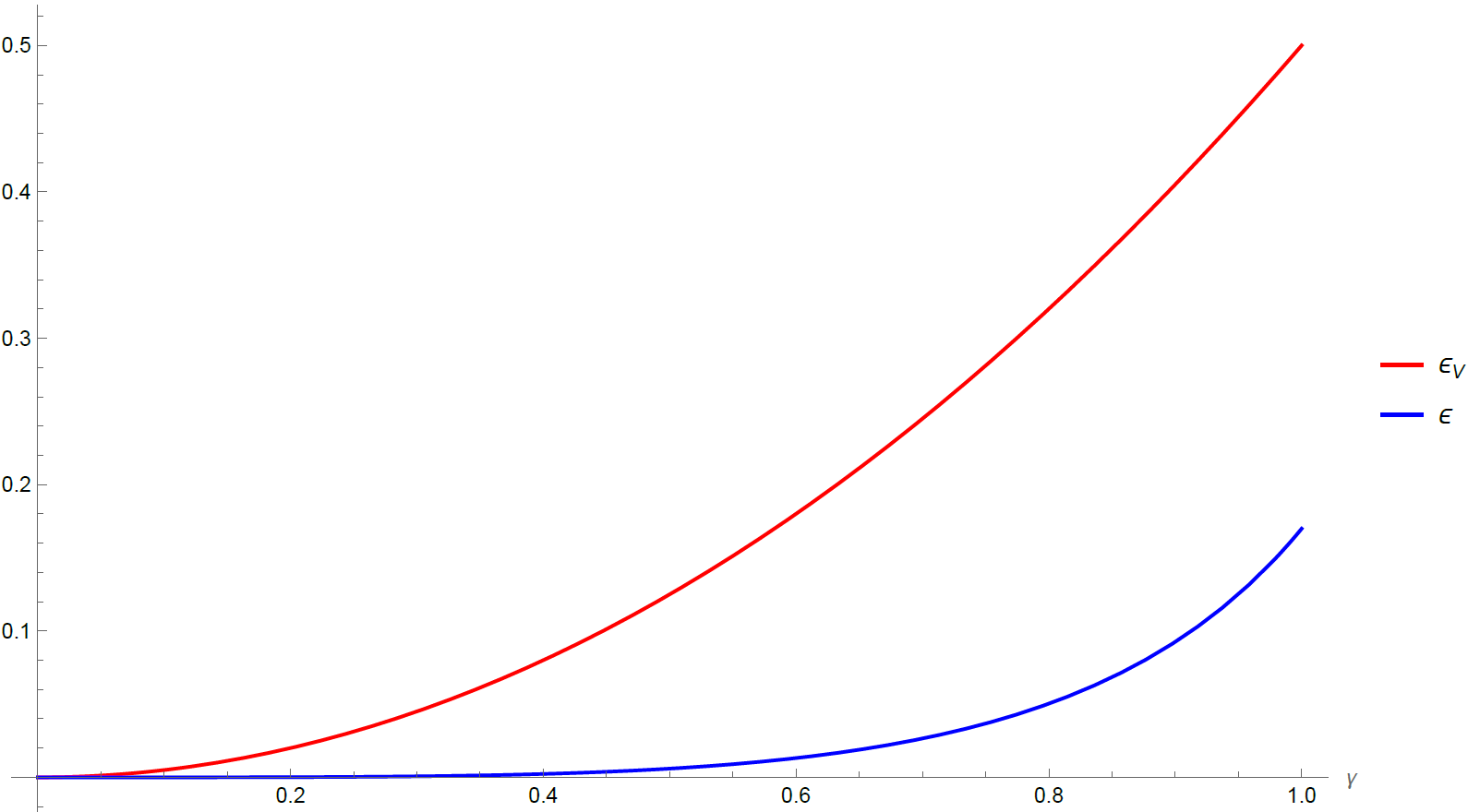}
		\caption{Parameters that assures the exponential inflation in units such that $\kappa=1$ for an exponential potential. The standard slow roll parameter is presented (red curve) and the Hubble parameter for the GUP scenario is also presented (blue curve). It is shown that $\epsilon<\epsilon_{V}$ in all the range. } \label{FigParameters}
	\end{figure}

Let us proceed in an analogous manner, in the Friedmann equation (\ref{Case1Friedmann}) we want to ignore the kinetic term $\frac{\dot{\phi}^2}{2}$ with respect to $F^2V$ and in the field equation (\ref{Case1ScalarEq}) we want to ignore $\ddot{\phi}$ with respect to the friction term $(3H-H_{F})\dot{\phi}$, with the hope that by doing so we will obtain an exponential expansion. Thus let us define the slow roll Hubble parameters as
	\begin{equation}\label{Case1DefHubbParameters}
		\epsilon=-\frac{\dot{H}}{H^2} , \hspace{1cm} \eta=\frac{\ddot{\phi}}{(3H-H_{F})\dot{\phi}} .
	\end{equation}
However in this case we obtain
	\begin{equation}
			\epsilon=\frac{3\dot{\phi}^2}{2F^{2}V}-\frac{3HH_{F}}{kF^2V} ,
	\end{equation}
thus, in this case assuming $\epsilon\ll1$ does not assure the first approximation i.e. it is not assured that the kinetic term can be dropped with respect to the potential, then we need to consider a third Hubble parameter directly defined as
	\begin{equation}\label{Case1DefHubbParameters2}
		\omega=\frac{\dot{\phi^2}}{2F^2V} ,
	\end{equation}
and thus the approximations will be justified with an approximately constant $H$ by demanding $\epsilon,\eta,\omega\ll1$. In this case however, these parameters can be related to the slow roll  parameters by
	\begin{equation}\label{Case1Epsilon}
		\epsilon=\epsilon_{V}\left[\frac{1+\frac{H_{F}}{3\epsilon_{V} H^2}\left(2H_{F}-3H-\frac{H^2_{F}}{3H}\right)}{1+\frac{H_{F}}{9H^2}\left(H_{F}-6H\right)}\right] ,
	\end{equation}
	\begin{equation}\label{Case1Eta}
		\eta=\frac{\eta_{V}}{3\left[1+\frac{H_{F}}{9H^2}(H_{F}-6H)\right]}-\frac{3H^2}{(3H-H_{F})^2}\epsilon-\frac{2H_{F}}{3H-H_{F}}-\frac{\dot{H}_{F}}{(3H-H_{F})^2} .
	\end{equation}
This opens a broader spectrum of possibilities since we don't require that the potential be restricted by the slow roll parameters, we only require that $\epsilon,\eta,\omega\ll1$, which can be done even if $\epsilon_{V}$ or $\eta_{V}$ are of order 1, the restrictions can be put in the form of the $F$ function.  However let us point out that the approximations used only assures us that in the Friedmann equation (\ref{Case1Friedmann}) we can ignore the kinetic term and thus obtain
	\begin{equation}\label{Case1AproxH}
		H^2\thickapprox \frac{\kappa}{3}F^2V .
	\end{equation} 
Therefore if we want to have an exponential expansion, we need to choose the $F$ function carefully. If we assume the standard slow roll parameters $\epsilon_{V},\eta_{V}\ll1$ we obtain that $V$ is approximately a constant, and thus if we consider $F$ as a function of $\phi$ only through its potential, that is $F=F(V(\phi))$ then the above analysis assures us that we will have an exponential expansion if $F$ is such that $\epsilon,\eta,\omega\ll1$. On the other hand, we note from (\ref{Case1AproxH}) that if $V$ can not be regarded as a constant, we can still have a constant Hubble parameter if $F$ is chosen in a correct way, that is if $F^2V$ is a constant. Let us consider a specific form to illustrate this point. In order to have a constant $\epsilon_{V}$ parameter  we choose $V=V_{0}e^{\gamma\phi}$, where $\gamma$ is a constant related to the GUP and $V_{0}$ is a constant, then in order to fulfil what we have described we choose
	\begin{equation}
		F=\sqrt{\frac{A}{V_{0}}}e^{-\frac{\gamma}{2}\phi} , 
	\end{equation}
where $H\thickapprox\frac{\kappa A}{3}$ with $A$ a positive constant, then we obtain
	\begin{equation}
		\epsilon\thickapprox\frac{\gamma^2}{2\kappa}\left[\frac{1-\sqrt{1-\frac{2\gamma^2}{3\kappa}}}{1-\frac{\gamma^2}{3\kappa}+\sqrt{1-\frac{2\gamma^2}{3\kappa}}}\right] ,
	\end{equation}
	\begin{equation}
		\eta\thickapprox\frac{\frac{2\gamma^2}{3\kappa}}{1-\frac{\gamma^2}{3\kappa}+\sqrt{1-\frac{2\gamma^2}{3\kappa}}}-\frac{\frac{2}{3}\epsilon}{1-\frac{\gamma^2}{3\kappa}+\sqrt{1-\frac{2\gamma^2}{3\kappa}}}-\frac{2\left(1-\sqrt{1-\frac{2\gamma^2}{3\kappa}}\right)}{1+\sqrt{1-\frac{2\gamma^2}{3\kappa}}} ,
	\end{equation} 
		\begin{equation}
		\omega\thickapprox\frac{3\kappa}{\gamma^2}\left[1-\frac{\gamma^2}{3\kappa}-\sqrt{1-\frac{2\gamma^2}{3\kappa}}\right].
	\end{equation}
Here the standard slow parameters are $\epsilon_{V}=\frac{\gamma^2}{2\kappa}$ and $\eta_{V}=\frac{\gamma^2}{\kappa}$. We note that the absolute value of the three parameters increase when $\gamma$ increases, therefore we can assure that they are small if $\gamma$ is small. In Figure \ref{FigParameters} we show the behaviour of the GUP Hubble parameter $\epsilon$ and the slow roll parameter $\epsilon_{V}$ (that coincides with the standard Hubble parameter). We note that $\epsilon<\epsilon_{V}$ in all the range and thus the case with the GUP turns out to be less restrictive that the standard slow roll scenario. For example choosing $\kappa=1$ and $\gamma=0.5$ we obtain $\epsilon\thickapprox0.00595$, $\eta\thickapprox-0.00217$, $\omega\thickapprox0.0455$ and thus the approximations are well justified whereas $\epsilon_{V}=0.125$ which is two orders of magnitude bigger than $\epsilon$.

So far we have considered the changes induced in the standard inflationary scenario by the modification of the Heisenberg algebra through approximations inspired by the standard slow roll scenario. However, it is possible to perform a full analytical analysis of the system of equations involved that will lead us to a very interesting result. The system of equations that describe the cosmological setup presented here is made of the hamiltonian constraint (\ref{SistStandardHC}) and the modified equations of motion (\ref{SistCase1-1}) and (\ref{SistCase1-2}) that we can express as
	\begin{equation}\label{Case1SisE2}
		P_{a}=-\frac{6a\dot{a}}{\kappa F}  , \hspace{1cm}
	P_{\phi}=\frac{a^3\dot{\phi}}{F} ,
	\end{equation}
	\begin{equation}\label{Case1SisE4}
		\dot{P}_{a}=-\left[\frac{\kappa}{12a^2}P^2_{a}-\frac{3}{2a^4}P^2_{\phi}+3a^2V\right]F , \hspace{1cm}
		\hspace{1cm} \dot{P}_{\phi}=-a^3V'F .
	\end{equation}
As we have said before $F$ has to depend on a parameter $\gamma$ such that $\lim_{\gamma\to0}F=1$ to recover the standard scenario. Let us propose the ansatz of an exponentially expanding universe, that is $a(t)=a_{0}e^{Ht}$ with $a_{0}$ and $H$ constants. It can be shown that the solution of this system of 5 equations without any approximation is
	\begin{equation}\label{Case1Solutions1}
		P_{a}=-\frac{6a^2_{0}H}{\kappa}\frac{e^{2Ht}}{F} , \hspace{1cm} P_{\phi}=\pm a^3_{0}e^{3Ht}\sqrt{\frac{2H\dot{F}}{\kappa F^3}} ,
	\end{equation}
	\begin{equation}\label{Case1Solutions2}
		\dot{\phi}=\pm\sqrt{\frac{2H}{\kappa}}\sqrt{\frac{\dot{F}}{F}} , \hspace{1cm}
		V=\frac{H}{\kappa F^3}\left(3HF-\dot{F}\right) .
	\end{equation}
Thus we neeed that $\frac{\dot{F}}{F}\geq0$. However, from the last expression we can express the Hubble parameter as
	\begin{equation}
		H=\frac{\dot{F}}{6F}\left[1\pm\sqrt{1+\frac{12\kappa F^4V}{\dot{F}^2}}\right] ,
	\end{equation}
then, since we want $H$ to be constant we can impose the condition $\dot{H}=0$ in the last expression, which leads us to 
	\begin{equation}
		\dot{F}^2-F\ddot{F}=0 ,
	\end{equation}
which has the  solution $F=Ae^{\alpha t}$ with $A$ and $\alpha$ constant. However, taking into account the behaviour with respect to $\gamma$ and for simplicity we choose $A=1$ and $\alpha$ a constant such that $\lim_{\gamma\to0}\alpha=0$. With this choice we obtain for the solution (\ref{Case1Solutions1}) and (\ref{Case1Solutions2})
	\begin{equation}\label{Case1SolutionsFinal}
		P_{a}=-\frac{6a^2_{0}H}{\kappa}e^{(2H-\alpha)t} , \hspace{1cm} P_{\phi}=\pm\sqrt{\frac{2H\alpha}{\kappa}}a^3_{0}e^{(3H-\alpha)t} , \hspace{1cm} \dot{\phi}=\pm\sqrt{\frac{2H\alpha}{\kappa}} ,
	\end{equation}
and the potential takes the form	
	\begin{equation}
		V=\frac{H}{\kappa}(3H-\alpha)e^{-2\alpha t} .
	\end{equation}
However now that we know the correct form of the potential, let us write it as 
	\begin{equation}\label{Case1Potential}
		V=Be^{-2\alpha t} ,
	\end{equation}
where $B$ will be an independent constant. Thus we obtain that the Hubble parameter will be described by two parameters, namely $\alpha$ and $B$ in the form
	\begin{equation}\label{Case1Hubble}
		H=\frac{\alpha}{6}\left(1+\sqrt{1+\frac{12\kappa B}{\alpha^2}}\right) .
	\end{equation}
Therefore by considering the form of the $F$ function to be an exponential in the time variable we can obtain a solution that describes an exponentially expanding universe with constant Hubble parameter without any approximation with an scalar field that will be linear in time and an exponential potential. Let us remember that $\alpha$ must depend on the constant $\gamma$ in such a way that $\lim_{\gamma\to0}\alpha=0$ and thus if $B$ is independent of $\gamma$ we note that
	\begin{equation}
		\lim_{\gamma\to0}H=\sqrt{\frac{\kappa B}{3}}.
	\end{equation}
We also can see that in this limit $\lim_{\gamma\to0}\dot{\phi}=0$ and $\lim_{\gamma\to0}V=B$ and thus we can identify $B$ in this limit with the cosmological constant, since in this limit the scalar field will vanish and the potential will only be a constant. Thus, the solution found is a generalization of the solution of a positive cosmological constant that is well known but now it appears in the form of an scalar field with its corresponding potential. We note from (\ref{Case1Hubble}) that since $\alpha\geq0$ the Hubble parameter is bigger than in the standard cosmological constant case and thus the expansion is faster in the presence of the GUP. Let us also remark that $B$ could be a function of $\gamma$, in that case if we consider for example that $\lim_{\gamma\to0}B=0$ then we obtain that $\lim_{\gamma\to0}V=0$, $\lim_{\gamma\to0}H=0$, that is we recover the static universe without expansion. Thus even if we start with an static universe, the presence of the GUP with the correct form of the $F$ function will create an exponentially expanding universe with an scalar field. Therefore the GUP is capable of creating and exponentially expanding universe in the form of an inflaton without the need of an extra source such as a cosmological constant.

We note that in our analysis we have not considered other sources for the gravitational field beyond the scalar field which turned out to depend on the $\gamma$ parameter describing the GUP. However although it is just a toy model for the universe, we can ask ourselves how to realize this solution in a more realistic scenario. For this let us remember that $\gamma$ is a parameter that should measure how great the modification to the standard Heisenberg algebra is and should be related to the scale in which we expect the effects of a quantum theory of gravity to be relevant. Thus we can think of this parameter as a function of the size of the universe or inversely of its energy density in such a way that in the very early universe we expect a non zero value for $\gamma$ (when the effects of the GUP are expected to be big) and as the expansion carries on $\gamma$ decreases such that for big values for the scale factor that correspond to classical values this parameter can be neglected and therefore the inflaton field vanishes as well as its corresponding potential in one case, or only the cosmological constant remains in the other. Therefore, we can have the standard cosmological scenario with the other sources of the universe and the scalar field does not decay, it just vanishes because the $\gamma$ parameter can be neglected.

Let us also remark an interesting point regarding the potential found. From (\ref{Case1Solutions1}) we can see that by neglecting the integration constant we obtain
	\begin{equation}\label{Case1ScalarField}
		\phi=\pm\sqrt{\frac{2H\alpha}{\kappa}}t , \hspace{1cm} V(\phi)=Be^{\mp\sqrt{\frac{2\kappa \alpha}{H}}\phi} .
	\end{equation}
Considering the minus sign for the field we obtain 
	\begin{equation}
		\frac{V'}{V}=\sqrt{\frac{12\kappa}{1+\sqrt{1+\frac{12B\kappa}{\alpha^2}}}} .
	\end{equation}
In the case in which $B$ is the cosmological constant and does not depend on $\gamma$ we have from the latter that $\lim_{\gamma\to0}\frac{V'}{V}=0$, thus we can say that in this case we are obtaining an inflationary scenario where the $\epsilon_{V}$ slow roll parameter (\ref{DefSlowRollP}) is satisfied when $\gamma\ll1$. However in the case of a static universe where $B$ depends on $\gamma$ we note that if we choose a dependence on $\gamma$ such that $\lim_{\gamma\to0}\frac{B}{\alpha^2}=0$  in units of $\kappa=1$ we will obtain $\lim_{\gamma\to0}\frac{V'}{V}=\sqrt{6}$ which is an order $1$ constant. Therefore in this case for $\gamma\to0$ we will have inflation but not fulfil the slow roll conditions. In fact we will be fulfilling the dS Swampland Conjecture \cite{Obied:2018sgi,Ooguri:2018wrx,Andriot:2018wzk,Agrawal:2018own,Roupec:2018mbn} which states that
	\begin{equation}\label{dSConjecture}
		\frac{V'}{V}\geq c ,
	\end{equation}
where $c$ is an order $1$ constant. In order to fulfil this conjecture in general we find the condition 
	\begin{equation}
		B\leq\frac{\alpha^2}{12}\left[\left(\frac{12}{c^2}-1\right)^2-1\right] ,
	\end{equation}
where  $c\in[1,\sqrt{6}]$. in particular if we choose for example $B=\alpha^2$ the latter simplifies to 	
	\begin{equation}
		1\leq\frac{1}{12}\left[\left(\frac{12}{c^2}-1\right)^2-1\right] ,
	\end{equation}
which can be fulfilled if $c\in[1,\sqrt{\sqrt{13}-1}]\thickapprox[1,1.614]$. Thus in the static universe case we have inflation fulfilling the dS conjecture for different values of the $c$ constant. This is an interesting result since the conjecture is expected to be fulfilled for theories that have a correct quantum behaviour and thus it is an agreement with the idea that the modification of the Heisenberg algebra that we are considering comes from the same type of argument.

To end this section let us remark that the condition to have a constant Hubble parameter was that $F$ has the form of an exponential with the time variable. However in this case $F$ must be a function of the coordinates. We note from the ansatz of the scale factor and (\ref{Case1ScalarField}) that in this case $F$ can be written as a power law of the scale factor or an exponential function of the scalar field in the form 
	\begin{equation}\label{Case1F}
		F=\left(\frac{a}{a_{0}}\right)^{c_{1}}=e^{c_{2}\phi} ,
	\end{equation}
where
	\begin{equation}
		c_{1}=\frac{\alpha}{H}=\frac{6}{1+\sqrt{1+\frac{12\kappa B}{\alpha^2}}} ,
	\end{equation}
	\begin{equation}
		c_{2}=\alpha\sqrt{\frac{\kappa}{2H\alpha}}=\sqrt{\frac{3\kappa}{1+\sqrt{1+\frac{12\kappa B}{\alpha^2}}}} .
	\end{equation}
Therefore by using the standard relation
	\begin{equation}
		\Delta x\Delta p\geq \frac{1}{2}|\langle[x,p]\rangle| ,
	\end{equation}
 the modified uncertainty principle will take the form
	\begin{equation}\label{Case1UPa}
		\Delta a\Delta P_{a} \geq \frac{1}{2a^{c_{1}}_{0}}|\langle a^{c_{1}}\rangle| ,
	\end{equation}
	\begin{equation}\label{Case1UPfi}
		\Delta\phi\Delta P_{\phi}\geq\frac{1}{2}|\langle e^{c_{2}\phi}\rangle| .
	\end{equation}
	
We note that as long as $B$ is independent of $\gamma$ or it depends on it in such a way that $\lim_{\gamma\to0}\frac{\alpha^2}{B}=0$, we will obtain  
		\begin{equation} \lim_{\gamma\to0}c_{1}=\lim_{\gamma\to0}c_{2}=0 .
		\end{equation} 
Therefore taking $\gamma\ll1$ we will always have $c_{1},c_{2}\ll1$. Thus, since $c_{2}$ is small we can expand the exponential keeping only up to quadratic terms. Then by choosing states in which $\langle\phi\rangle=0$ we have that
	\begin{equation}
		\langle e^{c_{2}\phi}\rangle \thickapprox \bigg\langle1+c_{2}\phi+\frac{c^2_{2}}{2}\phi^2 \bigg\rangle=1+\frac{c^2_{2}}{2}\langle\phi^2\rangle=1+\frac{c^2_{2}}{2}(\Delta\phi)^2.
	\end{equation}
Thus the uncertainty relations leads to
	\begin{equation}\label{Case1GUPSF}
		\Delta P_{\phi}\geq \frac{2+c^2_{2}(\Delta\phi)^2}{4\Delta\phi} ,
	\end{equation}
which has the same general form as the GUP that can be derived from the simplest form (\ref{SimplestGUP}) but applied to the momentum instead of the coordinate, that is, it is a particular form of the EUP. Therefore, we note that the right hand side has a minimum value in $\Delta\phi_{min}=\frac{\sqrt{2}}{c_{2}}$ which corresponds to a minimum measurable value for the scalar field momentum given by $(\Delta P_{\phi})_{min}=\frac{c_{2}}{\sqrt{2}}$. Thus, the first case leads to a minimum measurable momentum for the scalar field.

On the other hand, for the scale factor when we take $c_{2}\ll1$ we cannot reduce the right hand side of (\ref{Case1UPa}), thus we can not have a conclusion in general. However, we can  write down the expectation value in the right hand side of (\ref{Case1UPa}) as an indeterminate function of the scale factor uncertainty in the form
	\begin{equation}
		|\langle a^{c_{1}}\rangle|=I_{1}(\Delta a) ,
	\end{equation}
valid for at least some of the states considered. Therefore the uncertainty relation will lead to
	\begin{equation}\label{UncertaintyGeneralMomentumA}
		\Delta P_{a}\geq \frac{I_{1}(\Delta a)}{2a_{0}^{c_{1}}\Delta a} .
	\end{equation}
Thus, we will obtain a minimum measurable value for the momentum of the scale factor if the function $I_{1}(x)/x$ has a minimum. The specific form of the function will be determined by the value of the constant $c_{1}$, which is defined by the values of the Hubble constant $H$ during inflation and the GUP parameter $\alpha$ and the state in which the expectation value is calculated. There is however a particular choice that can lead to an strange behaviour, let us assume that we can take $c_{1}=2n$ with $n$ a natural number, this can be achieved if we take
	\begin{equation}
		B=\frac{\alpha^2}{12\kappa}\left[\left(\frac{3}{n}-1\right)^2-1\right] ,
	\end{equation}
which leads to $H=\frac{\alpha}{2n}$ and $c_{2}=\sqrt{\kappa n}$. In this case we note that $\lim_{\gamma\to0}B=0$, thus it is a particular case of the static universe. Furthermore, since $H$ depends linearly in $\alpha$ we note that in order to have a proper inflationary period we need $\alpha$ to be an order 1 constant, thus we are on an early universe scenario. In this case by using the property $\langle a^{2n}\rangle\geq(\langle a^{2}\rangle)^{n}$ and choosing states such that $\langle a \rangle=0$, the uncertainty relation (\ref{Case1UPa}) is simplified as
	\begin{equation}\label{Case1Uncertainty}
		\Delta P_{a}\geq \frac{(\Delta a)^{2n-1}}{2a_{0}^{2n}} .
	\end{equation}
This result has the same behaviour as (\ref{Case1GUPSF}) when $n=1$ for big values of $\Delta a$ that is the minimum value for $\Delta P_{a}$ grows linearly with $\Delta a$. However, for (\ref{Case1GUPSF}) there is a minimum value for the momentum uncertainty and if we consider smaller values for the coordinate uncertainty beyond this point, the uncertainty on the momentum increases as in the standard HUP scenario. On the other hand, in this result, the minimum value gets displaced to the origin and therefore our result leads to the unexpected behaviour that $\Delta P_{a}\to0$ implies $\Delta a\to0$ loosing the region of standard behaviour. Thus, we conclude that it is possible to obtain a minimum measurable value for the scale factor momentum as in the standard EUP case but we can also obtain some different phenomenology. The concrete result will depend on the specific value of the Hubble constant, the GUP parameter and the states under consideration.

\subsection{Case II}
\label{S-Case2}
Let us move on to consider GUPs that depend on the momenta. First of all we will consider the simplest case by using a generalization of the Heisenberg algebra independently for each coordinate defined in (\ref{DefCase2}). In this case the system of equations is defined by the hamiltonian constraint (\ref{SistStandardHC}) as well as the equations of motion
	\begin{equation}\label{Case2SisE2}
		P_{a}=-\frac{6a\dot{a}}{\kappa F_{1}}  ,\hspace{1cm}
		P_{\phi}=\frac{a^3\dot{\phi}}{F_{2}} ,
	\end{equation}
	\begin{equation}\label{Case2SisE4}
		\dot{P}_{a}=-\left[\frac{\kappa}{12a^2}P^2_{a}-\frac{3}{2a^4}P^2_{\phi}+3a^2V\right]F_{1} , \hspace{1cm}
		\hspace{1cm} \dot{P}_{\phi}=-a^3V'F_{2} .
	\end{equation}
with $F_{1}=F_{1}(a,P_{a})$ and $F_{2}=F_{2}(\phi,P_{\phi})$. The procedure to obtain the Friedmann equation will be the same as before that is, we need to specify the form in which the $F$ functions depend on the momenta such that we can solve for $P_{a}$ and $P_{\phi}$ in (\ref{Case2SisE2}) and then substitute these results back in (\ref{SistStandardHC}). As we haven seen in Section \ref{S-NoScalarField} this procedure will modify completely the form of the Friedmann equation. If we consider the simplest scenario where the modification is only made in the scale factor, that is if we take $F_{2}=1$, using this procedure we can obtain the Friedmann equations (\ref{ModifiedFriedmann1})-(\ref{ModifiedFriedmann3}) after considering the corresponding form for the $F_{1}$ function with the energy density $\rho$ replaced by $\frac{\dot{\phi}^2}{2}+V$ as expected. When $F_{2}$ is taken into account we can obtain more forms of the Friedmann equations. However instead of following this path that depends completely on the form of the $F$ functions we will perform the analytical treatment of the system of equations analogous to the one presented in the last section. In this case the system of equations can also be solved  with the ansatz $a=a_{0}e^{Ht}$ giving the solution
	\begin{equation}
		P_{a}=-\frac{6a^2_{0}H}{\kappa}\frac{e^{2Ht}}{F_{1}} , \hspace{1cm} P_{\phi}=\pm a^3_{0}e^{3Ht}\sqrt{\frac{2H\dot{F}_{1}}{\kappa F^3_{1}}} ,
	\end{equation}
	\begin{equation}\label{SolGeneralCampo}
		\frac{\dot{\phi}}{F_{2}}=\pm\sqrt{\frac{2H}{\kappa}}\sqrt{\frac{\dot{F}_{1}}{F^3_{1}}} , \hspace{1cm}
		V=\frac{H}{\kappa F^3_{1}}\left(3HF_{1}-\dot{F}_{1}\right) .
	\end{equation}  
We note that since we have considered separate functions for each coordinate, the only presence of the $F_{2}$ function is in the form of $\dot{\phi}$. In particular, the potential function has the same form as before but now with $F_{1}$, thus we can follow the same procedure as before, that is impose the condition $\dot{H}=0$ which leads us to $F_{1}=e^{\alpha t}$ once again. With this we obtain the correct form for the potential and thus we propose once again $V=Be^{-2\alpha t}$. The only difference with case I is that in this case the form of the scalar field is not determined since $F_{2}$ is not restricted by the system. Thus we still have the freedom to choose $F_{2}$ as we prefer and it will not alter the expansion of the universe, it will only alter the form of the scalar field and thus the form of the potential to realize this solution. Therefore with this solution we obtain
	\begin{equation}\label{Case2Soluciones}
		P_{a}=-\frac{6a^2_{0}H}{\kappa}e^{(2H-\alpha)t} , \hspace{1cm} P_{\phi}=\pm\sqrt{\frac{2H\alpha}{\kappa}}a^3_{0}e^{(3H-\alpha)t} , \hspace{1cm} \frac{\dot{\phi}}{F_{2}}=\pm\sqrt{\frac{2H\alpha}{\kappa}}e^{-\alpha t} ,
	\end{equation}
with the Hubble parameter given once again by
	\begin{equation}
		H=\frac{\alpha}{6}\left(1+\sqrt{1+\frac{12\kappa B}{\alpha^2}}\right) .
	\end{equation}
If we make a general factorization ansatz $F_{2}=f_{2}(P_{\phi})g_{2}(\phi)$ we obtain from the latter equation on (\ref{Case2Soluciones}) that the scalar field can be obtained by solving
\begin{equation}\label{Case2EcField}
	\int\frac{d\phi}{g_{2}(\phi)}=\pm\sqrt{\frac{2H\alpha}{\kappa}}\int f_{2}\left(\pm\sqrt{\frac{2H\alpha}{\kappa}}a^3_{0}e^{(3H-\alpha)t}\right)e^{-\alpha t}dt .
\end{equation}

Since the Hubble parameter and the potential (in terms of the time variable) has the same form as before the analysis presented in the last section for the Hubble parameter is applicable here as well. In particular $B$ can be identified with the cosmological constant or with a dependence on $\gamma$ that leads us to the static universe. What is new for the scale factor is that now the $F_{1}$ function can also be written as powers of the scale factor as in (\ref{Case1F}), but we note from (\ref{Case2Soluciones}) that we can also write it as powers of the scale factor momentum in the form
	\begin{equation}\label{Case2F1}
		F_{1}=\left(c_{3}P_{a}\right)^{c_{4}} ,
	\end{equation}
where
	\begin{equation}
		c_{3}=-\frac{\kappa}{6a^2_{0}H}=-\frac{\kappa}{a^2_{0}\alpha\left(1+\sqrt{1+\frac{12\kappa B}{\alpha^2}}\right)} ,
	\end{equation}
	\begin{equation}
		c_{4}=\frac{\alpha}{2H-\alpha}=\frac{6}{\sqrt{1+\frac{
		12\kappa B}{\alpha^2}}-5} .
	\end{equation}
Thus the uncertainty principle in this case takes the form
\begin{equation}
	\Delta a\Delta P_{a}\geq\frac{|c_{3}^{c_{4}}|}{2}|\langle P_{a}^{c_{4}}\rangle| .
\end{equation}
In the case in which $\lim_{\alpha\to0}\frac{\alpha^2}{B}=0$ we have that $c_{4}\ll1$ and  $c_{3}\to-\frac{1}{a^2_{0}}\sqrt{\frac{\kappa}{12B}}$ when $\gamma\to0$. Once again we can define an unknown function as
		\begin{equation}\label{UnknownFunction2}
			|\langle P_{a}^{c_{4}}\rangle|=I_{2}(\Delta P_{a}),
		\end{equation}
therefore the uncertainty principle will take the form
	\begin{equation}\label{UncertaintyRelationCoordinatelA}
		\Delta a\geq \frac{|c_{3}^{c_{4}}|I_{2}(\Delta P_{a}
				)}{2\Delta P_{a}} .
	\end{equation}
In this way, we can have a minimum measurable value of the scale factor if the function $I_{2}(x)/x$ has a minimum as in the standard GUP, or the same unexpected behaviour as in case I in which $\Delta a\to0$ implies $\Delta P_{a}\to 0$ if we can take $c_{4}=2n$ with $n$ a natural number. The specific result will depend once again on the values of the Hubble and GUP parameters as well as in the state in which the expectation value is computed.

On the other hand, since we have the freedom to choose $F_{2}$ in this case we can obtain the standard implications of the GUP by choosing it in a proper manner. For example, if we consider it to be a function of $P^2_{\phi}$ or a function of $\phi^2$ such that $F_{2}(x^2)/x$ has a minimum value we can obtain a minimal measurable value for the scalar field or for the momentum respectively. The only requirement to have an analytical expression for the potential in terms of the momentum is that we can solve (\ref{Case2EcField}) for $t$ since the form of $V$ with respect to time does not change. Let us consider two illustrative examples:
	\begin{itemize}
		\item Choosing $F_{2}=f_{2}(P_{\phi})$ we obtain from (\ref{Case2EcField}) that the scalar field will be given by
			\begin{equation}
				\phi=\mp\sqrt{\frac{2H}{\kappa\alpha}}\int f_{2}\left(\pm\sqrt{\frac{2H\alpha}{\kappa}}a^3_{0}x^{(1-3H/\alpha)}\right)dx ,
			\end{equation}
		where $x=e^{-2\alpha t}$ with the potential given by $V=Bx^2$, thus in order to write analytically $V=V(\phi)$ we need functions such that we can solve for $x$ in the latter expression.
		\item Choosing $F_{2}=g_{2}(\phi)$ we obtain that we can always solve for the time in terms of the scalar field and thus the potential will be given in general by
			\begin{equation}
				V(\phi)=\frac{3\kappa B}{1+\sqrt{1+\frac{12\kappa B}{\alpha^2}}}\left[\int\frac{d\phi}{g_{2}(\phi)}\right]^2 .
			\end{equation}
	\end{itemize}
Therefore, we conclude that in case II the phenomenology of the GUP in the scale factor can have a minimum measurable value for the momentum or for the coordinate, depending on the values of the Hubble constant, the GUP parameter and the chosen state. Whereas in the scalar field we have the freedom to choose the $F_{2}$ function in order to have the standard phenomenology of the GUP or the EUP.

Finally let us talk about a particular solution that has some interesting features. We note from (\ref{Case2Soluciones}) that if we choose $\alpha=2H$ then the scale factor momentum will be a constant. This will be helpful because we can make the separation $F_{1}(a,P_{a})=f_{1}(P_{a})g_{1}(a)=f_{1}(P_{a})e^{2Ht}$. Then $f_{1}(P_{a})$ will be constant independently of the form of $f_{1}$. Thus we can look for a solution to the system of equations (\ref{SistStandardHC}), (\ref{Case2SisE2}) and (\ref{Case2SisE4}) with the ansatz $V=Be^{-2\alpha t}=Be^{-4Ht}$. The solution obtained has the form
	\begin{equation}
		P_{a}=\pm6a^2_{0}\sqrt{\frac{B}{\kappa}} , \hspace{1cm} P_{\phi}=\pm2a^3_{0}e^{Ht}\sqrt{B} ,
	\end{equation}
and thus the scalar field is defined by
	\begin{equation}
		\frac{\dot{\phi}}{F_{2}}=\pm2\sqrt{B}e^{-2Ht} ,
	\end{equation}
and the Hubble parameter is given in terms of the $f_{1}$ function as
	\begin{equation}
		H=\mp\sqrt{\kappa B}f_{1}\left(\pm6a^2_{0}\sqrt{\frac{B}{\kappa}}\right) .
	\end{equation}
Thus we can propose forms of the $f_{1}$ function as the standard cases used in Section \ref{S-NoScalarField}. In particular if we propose $f_{1}(P_{a})=\frac{1}{1-\gamma P^2_{a}}$ we can obtain big values for $H$ by choosing properly $B$, in the neighbourhood of the value that makes $f_{1}$ divergent. We note that since $\lim_{\gamma\to0}H=\lim_{\gamma\to0}\alpha=0$ then $\lim_{\gamma\to0}B=0$ and thus this solution will describe an static universe in the mentioned limit.

\subsection{Case III}
\label{S-Case3}
Let us study now the case in which the function can depend on the momenta and it is the same for both coordinates defined by (\ref{DefCase3}). We have the same system of equations than in case I, but now $F$ is not just a function of the coordinates, it can depend on  the momenta as well. Thus we have the hamiltonian constraint (\ref{SistStandardHC}) and
	\begin{equation}\label{Case3SisE2}
		P_{a}=-\frac{6a\dot{a}}{\kappa F(a,\phi,P_{a},P_{\phi})}  , \hspace{1cm}
		P_{\phi}=\frac{a^3\dot{\phi}}{F(a,\phi,P_{a},P_{\phi})} ,
	\end{equation}
	\begin{equation}\label{Case3SisE4}
		\dot{P}_{a}=-\left[\frac{\kappa}{12a^2}P^2_{a}-\frac{3}{2a^4}P^2_{\phi}+3a^2V\right]F(a,\phi,P_{a},P_{\phi}) , 
		\hspace{0.5cm} \dot{P}_{\phi}=-a^3V'F(a,\phi,P_{a},P_{\phi}) .
	\end{equation}
Let us begin by considering the simpler case in which the dependence is only in the total momentum. However, we note from the hamiltonian constraint (\ref{SistStandardHC}) that our coordinate space is not defined by a  plane euclidean metric. It has a metric where the components are not constants and are not even positive definite that is we have for the metric in this space
	\begin{equation}
		G^{\mu\nu}=\begin{pmatrix}
			G^{aa} & G^{a\phi} \\
			G^{\phi a} & G^{\phi\phi}
		\end{pmatrix}=\begin{pmatrix}
			-\frac{\kappa}{12a} & 0 \\
			0 & \frac{1}{2a^3} 
		\end{pmatrix}.
	\end{equation}
Therefore the total momentum have to take into account this metric and thus it must be defined as the kinetic part of the hamiltonian constraint, that is 
	\begin{equation}\label{DefTotalMomentum}
		\mathcal{P}^2=-\frac{\kappa}{12a}P^2_{a}+\frac{1}{2a^3}P^2_{\phi}.
	\end{equation}
However with this definition we have from the hamiltonian constraint $\mathcal{P}^2=-a^3V$, and thus if $F=F(\mathcal{P}^2)$ we actually have $F=F(-a^3V(\phi))$, then  we effectively have the same scenario as in case I and all that was derived there is applicable to this case as well. In particular the Friedmann equation can be expressed in general analogously to (\ref{Case1Friedmann}) as
	\begin{equation}
		H^2=\frac{\kappa}{3}\left[\frac{\dot{\phi}^2}{2}+F^2(-a^3V(\phi))V(\phi)\right] .
	\end{equation}
The slow roll analysis can also be done in the same manner and the exact solution describing the exponentially expanding universe proceed in the same form, the only difference is that now the condition to obtain such solution is 
	\begin{equation}
		F(\mathcal{P}^2)=e^{\alpha t},
	\end{equation} 
but as we have found in that case $a=a_{0}e^{Ht}$ and $V(\phi)=Be^{-2\alpha t}$ thus for this solution
	\begin{equation}
		\mathcal{P}^2=-a^3V=-a^3_{0}Be^{(3H-2\alpha)t} ,
	\end{equation}
thus we can write $F$ as a power law of the total momentum as
	\begin{equation}
		F=\left(-\frac{\mathcal{P}^2}{a^3_{0}B}\right)^{\frac{\alpha}{3H-2\alpha}}=\left(-\frac{\mathcal{P}^2}{a^3_{0}B}\right)^{\frac{2}{\sqrt{1+\frac{12\kappa B}{\alpha^2}}-3}} .
	\end{equation}
However with this form it is not possible to extract general consequences to the uncertainty  relations of the coordinates.

Let us consider now the most general dependence on $F$ without the restriction of a dependence only through $\mathcal{P}^2$, that is let us consider $F=F(a,\phi,P_{a},P_{\phi})$. In this case the equations in (\ref{Case3SisE2}) represent a coupled system of equations that need to be solved for the momenta when we consider a particular form for the $F$ function and then substitute this solution back in the hamiltonian constraint in order to obtain the Friedmann equation. Therefore once we consider a dependence on the momenta we cannot obtain a general Friedmann equation. However it turns out that the system of equations have the same exact solution as in case I with the same condition, that is we require that $F=e^{\alpha t}$ and the same analysis follows obtaining in this way an exponentially expanding universe. In particular the solution takes the form (\ref{Case1SolutionsFinal}) with the potential (\ref{Case1Potential}) and the slow roll Hubble parameters is given by (\ref{Case1Hubble}). That is the scalar field is linear with time, the potential is exponential with the scalar field and both momenta are exponentials with time. Therefore we can write the $F$ functions as a function of the coordinates as in (\ref{Case1F}), or as a function of the scale factor momentum as in (\ref{Case2F1}) or now we can also write it as a function of the scalar field momentum as
	\begin{equation}\label{Case3F}
		F=(c_{5}P_{\phi})^{c_{6}} ,
	\end{equation}
where
	\begin{equation}
		c_{5}=\pm\frac{1}{a^3_{0}}\sqrt{\frac{\kappa}{2H\alpha}}=\pm\frac{1}{\alpha a^3_{0}}\sqrt{\frac{3\kappa}{1+\sqrt{1+\frac{12B\kappa}{\alpha^2}}}} , \hspace{1cm} c_{6}=\frac{\alpha}{3H-\alpha}=\frac{2}{\sqrt{1+\frac{12B\kappa}{\alpha^2}}-1}.
	\end{equation}
This leads to the following uncertainty relation
	\begin{equation}
		\Delta\phi\Delta P_{\phi}\geq \frac{|c_{5}^{c_{6}}|}{2}|\langle P_{\phi}^{c_{6}}\rangle| ,
	\end{equation}
where we note that $c_{6}$ is a small number when $\alpha\ll1$ but $c_{5}\gg1$ in the same limit. However $\lim_{\alpha\to0}c_{5}^{c_{6}}=1$ as expected. In the same way as in case II we write
	\begin{equation}
		|\langle P_{\phi}^{c_{6}}\rangle|=I_{3}(\Delta P_{\phi}) ,
	\end{equation}
with an unknown function $I_{3}$ determined by the value of $H$, $\alpha$ and the state under consideration, then the uncertainty relation takes the form
	\begin{equation}\label{UncertaintyUnknownMomentumField}
		\Delta\phi\geq \frac{|c_{5}^{c_{6}}|I_{3}(\Delta P_{\phi})}{2\Delta P_{\phi}} ,
	\end{equation}
and we will obtain a minimum measurable value for the scalar field if $I_{3}(x)/x$ has a minimum. Therefore, in case III we can obtain a minimum measurable value for the scalar field and its momentum if the Hubble and GUP parameters take the appropriate value and we choose an appropriate state. We note that in this case we cannot take $c_{6}=2n$ since that would imply that $\Delta\phi\to0$ implies $\Delta P_{\phi}\to0$ which is in contradiction to (\ref{Case1GUPSF}). Therefore, in this case we can avoid the unexpected behaviour.

\subsection{Case IV}
\label{S-Case4}

In the present subsection we will consider two cases in which the GUP is generalized to more variables in a non diagonal form defined by (\ref{DefCase4}).
\vskip .5truecm

\noindent
{\bf A. Case with $F_1=F_2$ and $G_1=G_2$}

Let us start with the simplest generalization of this kind that is let us consider $F_{1}=F_{2}=F$ and $G_{1}=G_{2}=G$ where $F$ and $G$ can depend on the coordinates and any of the momenta. We also need both functions to depend on the parameter $\gamma$ in such a way that in the limit $\gamma\to0$ we recover the standard commutators, therefore from the definition we require that
	\begin{equation}\label{Case4SLimits}
		\lim_{\gamma\to0}F=1 , \hspace{1cm} \lim_{\gamma\to0}G=0 .
	\end{equation}
With this simplification the system of equations is written as
	\begin{equation}\label{Case4SSist-1}
		\mathcal{H}=-\frac{\kappa P^2_{a}}{12a}+\frac{P^2_{\phi}}{2a^3}+a^3V=0 ,
	\end{equation} 
	\begin{equation}\label{Case4SSist-2}
		\dot{a}=-\frac{\kappa P_{a}}{6a}F+\frac{P_{\phi}}{a^3}G , \hspace{1cm}
		\dot{\phi}=-\frac{\kappa P_{a}}{6a}G+\frac{P_{\phi}}{a^3}F ,
	\end{equation}
	\begin{equation}\label{Case4SSist-4}
		\dot{P}_{a}=\left[-\frac{\kappa P^2_{a}}{12a^2}+\frac{3P^2_{\phi}}{2a^4}-3a^2V\right]F-a^3V'G ,
	\end{equation}
	\begin{equation}\label{Case4SSist-5}
		\dot{P}_{\phi}=\left[-\frac{\kappa P^2_{a}}{12a^2}+\frac{3P^2_{\phi}}{2a^4}-3a^2V\right]G-a^3V'F .
	\end{equation}
We recall that in cases I and III the system of equations gave us a general solution  where everything was written in terms of the function $F$ (\ref{Case1Solutions1}), (\ref{Case1Solutions2}) and we obtained a particular form for this function and thus a complete solution for the whole system after imposing the condition that $H$ must be a constant. However in this case we have the same number of equations and thus we expect that the system will not constrain $F$ and $G$ and we have to rely on giving appropriate ansatz for these functions in order to obtain a constant $H$. Let us propose the appropriate ansatz $a=a_{0}e^{Ht}$, then we can use equations (\ref{Case4SSist-1}) and (\ref{Case4SSist-2}) to write everything in terms of one of the momenta, let us choose $P_{\phi}$, then we obtain
	\begin{equation}\label{Case4SSolMomenta}
		P_{a}=\frac{6}{\kappa F}\left[\frac{P_{\phi}G}{a^2}-Ha^2\right] , \hspace{0.5cm}
		V=\left[\frac{3G^2}{\kappa F^2a^2}-\frac{1}{2}\right]\frac{P^2_{\phi}}{a^6}+\frac{1}{\kappa F^2}\left[3H^2-\frac{6HP_{\phi}G}{a^4}\right],
	\end{equation}
	\begin{equation}\label{Case4SSolMomentaF}
		\dot{\phi}=\frac{P_{\phi}}{a^3}\left(\frac{F^2-G^2}{F}\right)+\frac{HGa}{F} .
	\end{equation}
Thus from (\ref{Case4SSist-4}) and (\ref{Case4SSist-5}) we can obtain the equations
		\begin{equation}\label{Case4SDiffEq}
		-\frac{\kappa P^2_{a}}{3a^2}+\frac{3P^2_{\phi}}{a^4}=\frac{\dot{P}_{a}F-\dot{P}_{\phi}G}{F^2-G^2} , \hspace{1cm}
		V'=\frac{\dot{V}}{\dot{\phi}}=\frac{\dot{P}_{a}G-\dot{P}_{\phi}F}{a^3(F^2-G^2)} ,
	\end{equation}
which in general will lead us to a system of differential equations for $P_{\phi}$. So the analytical solution for the system will be obtained after solving such differential equations. However in this case both equations in (\ref{Case4SDiffEq}) lead to 
	\begin{multline}\label{Case4SMomentumEq}
		\frac{3(F^2-G^2)}{a^4}\left(1-\frac{4G^2}{\kappa F^2a^2}\right)P^2_{\phi}+\left(1-\frac{6}{\kappa a^2}\right)G\dot{P}_{\phi}+\frac{6G}{\kappa a^2}\left[6H-\frac{4HG^2}{F^2}+\frac{\dot{F}}{F}-\frac{\dot{G}}{G}\right]P_{\phi}\\+\frac{6Ha^2}{\kappa}\left[\frac{2HG^2}{F^2}-\frac{\dot{F}}{F}\right]=0 .
	\end{multline}
We note that this equation is hard to solve in general, therefore the procedure to obtain a solution and then impose $\dot{H}=0$ is hard to follow. However instead of finding a general constraint for $F$ and $G$ we can propose a convenient ansatz from the equations that we already have, let us note from (\ref{Case4SSist-2}) that we can write the Hubble parameter as
		\begin{equation}
		H=-\frac{\kappa P_{a}}{6a^2}F+\frac{P_{\phi}G}{a^4} ,
	\end{equation}	
and thus we can assure that $H$ is a constant if we propose the ansatz 
	\begin{equation}\label{Case4SAnsatzMomenta}
		P_{a}=-\frac{6Aa^2}{\kappa F} , \hspace{1cm} P_{\phi}=\frac{Ba^4}{G} .
	\end{equation}
where $A$ and $B$ are constants. Then (\ref{Case4SMomentumEq}) simplifies to
	\begin{multline}\label{Case4SMomentumEq2}
		\left[3B^2\left(1-\frac{F^2}{G^2}\right)+B\left(\frac{\dot{G}}{G}-4H\right)\right]a^2_{0}e^{2Ht}=\frac{12B^2}{\kappa}\left(\frac{G^2}{F^2}-1\right)\\+\frac{6}{\kappa}\left[2HB+\frac{2HG^2}{F^2}(H-2B)+\frac{\dot{F}}{F}(B-H)\right] .
	\end{multline}
Thus we can see that an exponential ansatz for both $F$ and $G$ will greatly simplify the above equation and as we saw in the last sections an exponential form for the $F$ function was indeed compatible with the system. However in this case we have to take into account (\ref{Case4SLimits}), thus we propose
	\begin{equation}
		F=e^{\alpha t} , \hspace{1cm} G(t)=\delta e^{\beta t} ,
	\end{equation}
where $\alpha,\beta$ and $\delta$ are constants depending on $\gamma$ in such a way that $\lim_{\gamma\to0}\alpha,\beta,\delta=0$. Then (\ref{Case4SMomentumEq2}) simplifies to
	\begin{multline}
		\left[3B+\beta-4H\right]Ba^2_{0}e^{2Ht}-\frac{3B^2}{\delta^2}a^2_{0}e^{2(H+\alpha-\beta)t}-\frac{12\delta^2}{\kappa}\left[B^2+H(H-2B)\right]e^{2(\beta-\alpha)t}\\+\frac{6}{\kappa}\left[2B^2-2HB+\alpha(H-B)\right]=0 .
	\end{multline}
The consistent solution is obtained when $\beta=\alpha$ and leads to the following system of equations
	\begin{equation}\label{Case4SConstSys1}
		3B=[3B+\alpha-4H]\delta^2 ,
	\end{equation}
	\begin{equation}\label{Case4SConstSys2}
		2B^2-2HB+\alpha(H-B)=2\delta^2[B^2+H(H-2B)] .
	\end{equation}
With this ansatz the potential in (\ref{Case4SSolMomenta}) takes the form
		\begin{equation}
			V=\frac{3(H-B)^2}{\kappa}e^{-2\alpha t}-\frac{B^2a^2_{0}}{2\delta^2}e^{2(H-\alpha)t} .
		\end{equation}
However, as we have seen before it is convenient to keep an independent constant arriving from the potential since it will have a particular physical meaning, thus we propose the ansatz
	\begin{equation}
		V=De^{-2\alpha t}+Ee^{2(H-\alpha)t} ,
	\end{equation}
where $D$ will be an independent constant but $E$ will be written in terms of the others. Then we have the system of equations (\ref{Case4SConstSys1}), (\ref{Case4SConstSys2}) and from the above $D=\frac{3(H-B)^2}{\kappa}$. Thus we can write everything in terms of two variables, it is convenient to choose $D$ and $\delta$ as these two independent constants and thus we obtain the solution
	\begin{equation}\label{Case4ConsSolutions1}
		B=2\sqrt{\frac{\kappa D}{3}}\left[\frac{\delta^2(\delta^2-2)}{3-\delta^2}\right] , \hspace{1cm} \alpha=2\sqrt{\frac{\kappa D}{3}}\left[\frac{\delta^2(\delta^2-1)}{3-\delta^2}\right] ,
	\end{equation}
	\begin{equation}\label{Case4ConsSolutions2}
		H=\sqrt{\frac{\kappa D}{3}}\left[\frac{2\delta^4-5\delta^2+3}{3-\delta^2}\right] .
	\end{equation}
Then since $\lim_{\gamma\to0}\delta=0$ we have $\lim_{\gamma\to0}\alpha=0$ as expected but we also have $\lim_{\gamma\to0}B=0$ and $\lim_{\gamma\to0}H=\sqrt{\frac{\kappa D}{3}}$. Consequently if $D$ does not depend on $\gamma$ we can identify it once again with the cosmological constant, but if $\lim_{\gamma\to0}D=0$ then we will have the static universe case in this scenario. Then in this particular non-diagonal case we have the same physics as in the previous sections.

Since we want a positive value for the Hubble parameter from (\ref{Case4ConsSolutions2}) $\delta$ is restricted to the regions $\delta^2<1$ or $\frac{3}{2}<\delta^2<3$. Whereas if we want the expansion to be faster than in the cosmological constant case it is restricted to the region $2<\delta^2<3$. With this solution from (\ref{Case4SSolMomentaF}) we have that setting the integration constant to zero the scalar field takes the form
	\begin{equation}\label{Case4SScalarField}
		\phi=\frac{\delta(\delta^2-1)a_{0}}{2\delta^4-5\delta^2+3}e^{Ht} ,
	\end{equation}
and thus in the limit $\delta\to0$ the kinetic term vanishes as in the previous cases. The potential takes the form of two power terms in the field as
	\begin{equation}\label{Case4SPotential}
		V=D\left[\left(\frac{2\delta^4-5\delta^2+3}{\delta(\delta^2-1)a^{2}_{0}}\phi\right)^{\frac{4\delta^2(1-\delta^2)}{2\delta^4-5\delta^2+3}}-\frac{2a^2_{0}\kappa \delta^2(\delta^2-2)^2}{3(3-\delta^2)^2}\left(\frac{2\delta^4-5\delta^2+3}{\delta(\delta^2-1)a^{2}_{0}}\phi\right)^{\frac{6(1-\delta^2)}{2\delta^4-5\delta^2+3}}\right] ,
	\end{equation}
and obeys $\lim_{\gamma\to0}V=D$ as expected. therefore, we note from (\ref{Case4SScalarField}) and (\ref{Case4SAnsatzMomenta}) that in this case both momenta, the scale factor and the scalar field are exponential functions on the time variable, and thus the $F$ and $G$ functions can be written as powers of any of the four variables. We have that
	\begin{equation}
		F=\left(\frac{a}{a_{0}}\right)^{d_{2}}=(d_{1}\phi)^{d_{2}}=(d_{3}P_{a})^{d_{4}}=(d_{5}P_{\phi})^{d_{6}} ,
	\end{equation}
where
	\begin{equation}
		d_{1}=\frac{2\delta^4-5\delta^2+3}{\delta(\delta^2-1)a_{0}} , \hspace{1cm}  d_{2}=\frac{2\delta^2(\delta^2-1)}{2\delta^4-5\delta^2+3}, \hspace{1cm} 
		d_{3}=-\frac{1}{2a^2_{0}}\sqrt{\frac{\kappa}{3D}} , \hspace{1cm}
	\end{equation}
	\begin{equation} d_{4}=\frac{\delta^2}{\delta^2-3} , \hspace{1cm}
		d_{5}=\sqrt{\frac{3}{\kappa D}}\frac{3-\delta^2}{2a^4_{0}\delta(\delta^2-2)} , \hspace{1cm} d_{6}=\frac{\delta^2}{3(\delta^2-2)}
	\end{equation}
and $G=\delta F$. Thus the uncertainty relations lead to
	\begin{equation}\label{Case3UR1}
		\Delta a\Delta P_{a}\geq \frac{1}{2a^{d_{1}}_{0}}|\langle a^{d_{1}}\rangle|=\frac{|d_{3}^{d_{4}}|}{2}|\langle P_{a}^{d_{4}}\rangle|,
	\end{equation}
	\begin{equation}\label{Case3UR2}
		\Delta\phi\Delta P_{\phi}\geq\frac{|d_{2}^{d_{1}}|}{2}|\langle\phi^{d_{1}}\rangle|=\frac{|d_{5}^{d_{6}}|}{2}|\langle P_{\phi}^{d_{6}}\rangle | ,
	\end{equation}
	\begin{equation}\label{Case3UR3}
		\Delta a\Delta P_{\phi}\geq\frac{\delta}{2a^{d_{1}}_{0}}|\langle a^{d_{1}}\rangle |=\frac{\delta|d_{5}^{d_{6}}|}{2}|\langle P_{\phi}^{d_{6}}\rangle | ,
	\end{equation}
	\begin{equation}\label{Case3UR4}
		\Delta\phi\Delta P_{a}\geq\frac{\delta|d_{2}^{d_{1}}|}{2}|\langle \phi^{d_{1}}\rangle |=\frac{\delta|d_{3}^{d_{4}}|}{2}|\langle P_{a}^{d_{4}}\rangle| .
	\end{equation}
Since $\lim_{\gamma\to0}\delta=0$ we expect that $\delta\ll1$. Thus $d_{2}$ is a small number, whereas $d_{1}$ and $d_{5}$ are big numbers. In this case however we cannot propose $d_{2}, d_{4}$ or $d_{6}$ to be odd integers, it can only be done in the region $\frac{3}{2}<\delta^2<2$ and if chosen that way we would obtained the same unexpected behaviour discussed in the previous cases, that is that all the uncertainties on the coordinates and the momenta will vanish at the same time. However, we can write (\ref{Case3UR1}) as in (\ref{UncertaintyGeneralMomentumA}) or in (\ref{UncertaintyRelationCoordinatelA}) by using appropriate unknown functions defined by the values of the GUP parameter and the specific state considered. In the same way we can write (\ref{Case3UR2}) as in (\ref{UncertaintyUnknownMomentumField}) or an expression in the form of (\ref{UncertaintyGeneralMomentumA}) for the scalar field momentum. Thus, we can have minimum measurable values for both coordinates and momenta if the GUP parameter takes the values  such that the functions defined in this way allow it for some states. The same forms can be derived from (\ref{Case3UR3}) and (\ref{Case3UR4}) but relating the uncertainty of a coordinate with the uncertainty in the momentum of the other coordinate.

Let us note that this solution was obtained through a convenient ansatz on the momenta that assure us that the Hubble parameter will be constant. In this case we did not find a general constraint in $F$ and $G$ for this to happen and thus we can not assure that it is the only way to obtain a constant Hubble parameter.

\vskip .5truecm

\noindent
{\bf B. Case with $F_1\not=F_2$ and $G_1\not=G_2$}

Finally let us study the most general case in which $F_{1}\neq F_{2}$ and $G_{1}\neq G_{2}$ defined by (\ref{DefCase4}). These functions must obey
	\begin{equation}\label{Case4Limits}
		\lim_{\gamma\to0}F_{1}=\lim_{\gamma\to0}F_{2}=1 , \hspace{1cm} \lim_{\gamma\to0}G_{1}=\lim_{\gamma\to0}G_{2}=0.
	\end{equation}
In this case the system of equation reads
	\begin{equation}\label{Case4Sist1}
		\mathcal{H}=-\frac{\kappa P^2_{a}}{12a}+\frac{P^2_{\phi}}{2a^3}+a^3V=0 ,
	\end{equation} 
	\begin{equation}\label{Case4Sist2}
		\dot{a}=-\frac{\kappa P_{a}}{6a}F_{1}+\frac{P_{\phi}}{a^3}G_{1} ,\hspace{1cm}
		\dot{\phi}=-\frac{\kappa P_{a}}{6a}G_{2}+\frac{P_{\phi}}{a^3}F_{2},
	\end{equation}
	\begin{equation}\label{Case4Sist4}
		\dot{P}_{a}=\left[-\frac{\kappa P^2_{a}}{12a^2}+\frac{3P^2_{\phi}}{2a^4}-3a^2V\right]F_{1}-a^3V'G_{1} ,
	\end{equation}
	\begin{equation}\label{Case4Sist5}
		\dot{P}_{\phi}=\left[-\frac{\kappa P^2_{a}}{12a^2}+\frac{3P^2_{\phi}}{2a^4}-3a^2V\right]G_{2}-a^3V'F_{2} .
	\end{equation}	
We proceed as in the particular case studied before, that is we use (\ref{Case4Sist1}) and (\ref{Case4Sist2}) to write everything in terms of one momentum obtaining
	\begin{equation}
		P_{a}=\frac{6}{\kappa F_{1}}\left[\frac{P_{\phi}G_{1}}{a^2}-a^2H\right] , \hspace{1cm} 
		\dot{\phi}=\frac{P_{\phi}}{F_{1}a^3}\left(F_{1}F_{2}-G_{1}G_{2}\right)+\frac{aHG_{2}}{F_{1}} ,
	\end{equation}
	\begin{equation}
		V=\left[\frac{3G^2_{1}}{\kappa F^2_{1}a^2}-\frac{1}{2}\right]\frac{P^2_{\phi}}{a^6}+\frac{3}{\kappa F^2_{1}}\left[H^2-\frac{2HG_{1}P_{\phi}}{a^4}\right] .
	\end{equation}
Then we use (\ref{Case4Sist4}) and (\ref{Case4Sist5}) to obtain 
	\begin{equation}\label{Case4EcMomentum1}
		\left(1-\frac{4G^2_{1}}{\kappa F^2_{1}a^2}\right)\frac{3P^2_{\phi}}{a^4}-\frac{12}{\kappa F^2_{1}}\left(H^2a^2-\frac{2HG_{1}}{a^2}P_{\phi}\right)=\frac{\dot{P}_{a}F_{2}-\dot{P}_{\phi}G_{1}}{F_{1}F_{2}-G_{1}G_{2}} ,
	\end{equation}
	\begin{equation}\label{Case4EcMomentum2}
		V'=\frac{\dot{P}_{a}G_{2}-\dot{P}_{\phi}F_{1}}{a^3(F_{1}F_{2}-G_{1}G_{2})} .
	\end{equation}
In the particular case we only obtained one differential equation, however in this case these two expressions will lead to two different equations for $P_{\phi}$, from (\ref{Case4EcMomentum1}) we obtain
	\begin{multline}\label{Case4EcMomentumC1}
		\frac{3(F_{1}F_{2}-G_{1}G_{2})}{a^4}\left(1-\frac{4G_{1}^2}{\kappa F^2_{1}a^2}\right)P^2_{\phi}+\left(1-\frac{6F_{2}}{\kappa a^2F_{1}}\right)G_{1}\dot{P}_{\phi}\\+\frac{6F_{2}}{\kappa a^2F_{1}}\left[G_{1}\left(6H-\frac{4HG_{1}G_{2}}{F_{1}F_{2}}+\frac{\dot{F_{1}}}{F_{1}}\right)-\dot{G}_{1}\right]P_{\phi}+\frac{6Ha^2}{\kappa F^2_{1}}\left[2HG_{1}G_{2}-\dot{F_{1}}F_{2}\right]=0.	
	\end{multline}
Whereas from (\ref{Case4EcMomentum2}) we obtain
	\begin{multline}\label{Case4EcMomentumC2}
		\frac{6(F_{1}F_{2}-G_{1}G_{2})}{Ha^4}\left[\frac{G_{1}-G_{2}}{\kappa F^2_{1}a^2}\left(\dot{G}_{1}-\frac{G_{1}\dot{F}_{1}}{F_{1}}\right)+\frac{2HG_{1}}{\kappa F^2_{1}a^2}(G_{2}-2G_{1})+\frac{H}{2}\right]P^2_{\phi}\\+\frac{6G_{1}(G_{1}-G_{2})}{\kappa HF^2_{1}a^6}(F_{1}F_{2}-G_{1}G_{2})P_{\phi}\dot{P}_{\phi}+G_{2}\left[1+\frac{6}{\kappa a^2}\left(\frac{G_{1}}{F^2_{1}}(G_{1}-G_{2})-\frac{F_{2}G_{1}}{F_{1}G_{2}}\right)\right]\dot{P}_{\phi}\\+\frac{6}{\kappa F^2_{1}a^2}\left[\dot{F}_{1}\left(F_{2}(2G_{1}-G_{2})+\frac{2G_{1}G_{2}}{F_{1}}(G_{2}-G_{1})\right)-\dot{G}_{1}\left(F_{1}F_{2}+G_{2}(G_{2}-G_{1})\right)\right.\\\left.+2H\left(F_{1}F_{2}(2G_{1}+G_{2})-2G^2_{1}G_{2}\right)\right]P_{\phi}-\frac{6Ha^2}{\kappa F^2_{1}}\left[\frac{\dot{F}_{1}}{F_{1}}G_{2}(G_{2}-G_{1})+F_{2}\dot{F}_{1}-2HG^2_{2}\right]=0 .
	\end{multline}
Once again we note from (\ref{Case4Sist2}) that the Hubble parameter can be written as
	\begin{equation}
		H=-\frac{\kappa P_{a}}{6a^2}F_{1}+\frac{P_{\phi}}{a^4}G_{1}.
	\end{equation}
Thus in order to always have a constant parameter we propose
	\begin{equation}
		P_{a}=-\frac{6Aa^2}{\kappa F_{1}} , \hspace{1cm} P_{\phi}=\frac{Ba^4}{G_{1}}.
	\end{equation}
Consequently the system of equations (\ref{Case4EcMomentumC1}) and (\ref{Case4EcMomentumC2}) simplifies to
	\begin{multline}
		B\left[3B\left(\frac{F_{1}F_{2}}{G^2_{1}}-\frac{G_{2}}{G_{1}}\right)+4H-\frac{\dot{G}_{1}}{G_{1}}\right]a^2=\frac{6H}{\kappa}\left(\frac{F_{2}\dot{F}_{1}}{F^2_{1}}-\frac{2HG_{1}G_{2}}{F^2_{1}}\right)\\+\frac{12B^2}{\kappa}\left(\frac{F_{2}}{F_{1}}-\frac{G_{1}G_{2}}{F^2_{1}}\right)+\frac{6BF_{2}}{\kappa F_{1}}\left(\frac{4HG_{1}G_{2}}{F_{1}F_{2}}-2H-\frac{\dot{F}_{1}}{F_{1}}\right) ,
	\end{multline}
	\begin{multline}
		\frac{B}{G_{1}}\left[3B\left(\frac{F_{1}F_{2}}{G_{1}}-G_{2}\right)+G_{2}\left(4H-\frac{\dot{G}_{1}}{G_{1}}\right)\right]a^2\\=\frac{6B^2}{\kappa H}\left[\left(1-\frac{G_{2}}{G_{1}}\right)\left(\frac{G_{1}G_{2}}{F^2_{1}}-\frac{F_{2}}{F_{1}}\right)\left(4H-\frac{\dot{G}_{1}}{G_{1}}\right)\right.\\\left.+ \left(\frac{G_{2}}{F^2_{1}G_{1}}-\frac{F_{2}}{F_{1}G^2_{1}}\right)\left((G_{1}-G_{2})\left(\dot{G}_{1}-\frac{G_{1}\dot{F}_{1}}{F_{1}}\right)+2HG_{2}(G_{2}-2G_{1})\right)\right]\\+\frac{6B}{\kappa}\left\{G_{2}\left(\frac{F_{2}}{F_{1}G_{2}}-\frac{G_{1}-G_{2}}{F^2_{1}}\right)\left(4H-\frac{\dot{G}_{1}}{G_{1}}\right)-\frac{1}{G_{1}F^2_{1}}\left[-\dot{G}_{1}(F_{1}F_{2}+G_{2}(G_{2}-G_{1}))\right.\right.\\\left.\left.+\dot{F}_{1}\left(F_{2}(2G_{1}-G_{2})+\frac{2G_{1}G_{2}}{F_{1}}(G_{2}-G_{1})\right)+2H(F_{1}F_{2}(2G_{1}+G_{2})-2G^2_{1}G_{2})\right]\right\}\\+\frac{6H}{\kappa F^2_{1}}\left[\frac{\dot{F}_{1}}{F_{1}}G_{2}(G_{2}-G_{1})+F_{2}\dot{F}_{1}-2HG^2_{2}\right] .
	\end{multline}
From both equations it can be seen that a solution will be obtained if we propose an exponential ansatz for the 4 functions, however we have to take into account the limiting behaviours (\ref{Case4Limits}), thus we propose
	\begin{equation}
		F_{1}=e^{\alpha t} , \hspace{1cm} F_{2}=Ce^{\alpha t} , \hspace{1cm} G_{1}=h_{1}e^{\alpha t}, \hspace{1cm} G_{2}=h_{2}e^{\alpha t} ,
	\end{equation}
where $\lim_{\gamma\to0}\alpha,h_{1,2}=0$ and $\lim_{\gamma\to0}C=1$.  With this ansatz we obtain $h_{1}=h_{2}=\eta$ and the system of equations
	\begin{equation}\label{AuxAux1}
		\frac{3B}{\eta}\left(\frac{C}{\eta}-\eta\right)+4H-\alpha=0 ,
	\end{equation}
		\begin{equation}
		2B^2-2HB+\alpha(H-B)=\frac{2\eta^2}{C}\left[B^2-H(H-2B)\right] .
	\end{equation}
Then we are lead once again to a form of the potential consisting of two exponentials with time, and then we propose
	\begin{equation}\label{Case4Potential}
		V=De^{-2\alpha t}+Ee^{2(H-\alpha)t} ,
	\end{equation}
where $D$ is an independent constant related to $B$ and $H$ by $D=\frac{3(H-B)^2}{\kappa}$ and $E$ will depend on the other constants. Then solving the system of equations we obtain the solution
	\begin{equation}
		B=2\sqrt{\frac{\kappa D}{3}}\frac{\eta^2(\eta^2-2C)}{C(3C-\eta^2)} , \hspace{1cm} \alpha=2\sqrt{\frac{\kappa D}{3}}\frac{\eta^2(\eta^2-C)}{C(3C-\eta^2)} ,
	\end{equation}
	\begin{equation}
		H=\sqrt{\frac{\kappa D}{3}}\left[\frac{2\eta^4-5\eta^2C+3C^2}{C(3C-\eta^2)}\right] ,
	\end{equation}
and we also have
	\begin{equation}\label{Case4KineticField}
		\dot{\phi}=a_{0}\sqrt{\frac{\kappa D}{3}}\frac{\eta(\eta^2-C)}{3C-\eta^2}e^{Ht} .
	\end{equation}
We know that $\lim_{\gamma\to0}\eta=0$ and thus $\lim_{\gamma\to0}\alpha=\lim_{\gamma\to0}B=\lim_{\gamma\to0}\dot{\phi}=0$ and $\lim_{\gamma\to0}H=\sqrt{\frac{\kappa D}{3}}$ with $\lim_{\gamma\to0}V=D$. Thus once again we can identify $D$ with the cosmological constant if it does not depend on $\gamma$, but even if it depends on $\gamma$ and vanishes in that limit leading to a static universe we still have an exponential expansion thanks to the GUP. We require that $H\geq0$ which in this case will constrain the possible values of $\eta$ and $C$ and thus it will be easier to fulfil because of the presence of $C$. We also note that we can have a faster expansion than in the cosmological constant case by also constricting these two parameters leading to $2C<\eta^2<3C$, which can be less restrictive to $\eta$ than in the particular case by choosing appropriate values of $C$. Integrating (\ref{Case4KineticField}) and setting the integration constant to vanish we obtain the scalar field
	\begin{equation}
		\phi=\frac{a_{0}C\eta(\eta^2-C)}{2\eta^4-5\eta^2C+3C^2}e^{Ht} ,
	\end{equation}
and thus the potential takes the form of two power terms once again
	\begin{multline}
		V=D\left[\left(\frac{2\eta^4-5\eta^2C+3C^2}{a_{0}C\eta(\eta^2-C)}\phi\right)^{-\frac{4\eta^2(\eta^2-C)}{2\eta^4-5\eta^2C+3C^2}}\right.\\\left.-\frac{2\kappa a^2_{0}\eta^2(\eta^2-2C)^2}{3C^2(3C-\eta^2)^2}\left(\frac{2\eta^4-5\eta^2C+3C^2}{a_{0}C\eta(\eta^2-C)}\phi\right)^{-\frac{6C(\eta^2-C)}{2\eta^4-5\eta^2C+3C^2}}\right] .
	\end{multline}
Finally, since in this case we obtain that both the momenta and the coordinates are exponentials in time, we can write the $F$'s and $G$'s functions as powers of any of those, then we have that
	\begin{equation}
		F_{1}=\left(\frac{a}{a_{0}}\right)^{k_{2}}=(k_{1}\phi)^{k_{2}}=(k_{3}P_{a})^{k_{4}}=(k_{5}P_{\phi})^{k_{6}} ,
	\end{equation}
where
	\begin{equation}
		k_{1}=\frac{2\eta^4-5\eta^2C+3C^2}{\eta(\eta^2-C)a_{0}C} , \hspace{1cm}  k_{2}=\frac{2\eta^2(\eta^2-C)}{2\eta^4-5\eta^2C+3C^2} , \hspace{1cm} 
		k_{3}=-\frac{1}{2a^2_{0}}\sqrt{\frac{\kappa}{3D}} , \hspace{1cm}
	\end{equation}
	\begin{equation} 
		k_{4}=\frac{\eta^2}{\eta^2-3C} , \hspace{1cm}
		k_{5}=\sqrt{\frac{3}{\kappa D}}\frac{C(3C-\eta^2)}{2a^4_{0}\eta(\eta^2-2C)} , \hspace{1cm} k_{6}=\frac{\eta^2}{3(\eta^2-2C)} ,
	\end{equation}
and $F_{2}=CF_{1}$, $G_{1}=G_{2}=\eta F_{1}$. Thus the uncertainty relations lead to
	\begin{equation}
		\Delta a\Delta P_{a}\geq \frac{1}{2a^{k_{1}}_{0}}|\langle a^{k_{1}}\rangle |=\frac{|k_{3}^{k_{4}}|}{2}|\langle P_{a}^{k_{4}}\rangle |,
	\end{equation}
	\begin{equation}
		\Delta\phi\Delta P_{\phi}\geq\frac{|Ck_{2}^{k_{1}}|}{2}|\langle \phi^{k_{1}} \rangle |=\frac{|Ck_{5}^{k_{6}}|}{2}|\langle P_{\phi}^{k_{6}}\rangle | ,
	\end{equation}
	\begin{equation}
		\Delta a\Delta P_{\phi}\geq\frac{h}{2a^{k_{1}}_{0}}|\langle a^{k_{1}} \rangle |=\frac{h|k_{5}^{k_{6}}|}{2}|\langle P_{\phi}^{k_{6}}\rangle | ,
	\end{equation}
	\begin{equation}
		\Delta\phi\Delta P_{a}\geq\frac{h|k_{2}^{k_{1}}|}{2}|\langle \phi^{k_{1}} \rangle |=\frac{h|k_{3}^{k_{4}}|}{2}|\langle P_{a}^{k_{4}}\rangle | .
	\end{equation}
We note that these results have the same general form as the ones obtained in case A. Therefore, we conclude that we can define unknown functions to simplify the uncertainty relations in such a way that it is possible to obtain minimum measurable values for the coordinates and their momenta depending in this case on the GUP parameters $\delta$ and $C$ and the state under consideration. Furthermore, once again we can not make $k_{2}, k_{4}$ or $k_{6}$ to be an odd integer consistently with the limits of $\eta$ and $C$ when $\gamma\to0$ and thus we can avoid the unexpected behaviours.

\section{Final Remarks}
\label{S-FinalRemarks}

In the present article we have presented a general treatment of the modifications to the Friedmann and scalar field equations when we take into account a GUP by using a classical limit and study the inflationary scenarios within this framework. This limit was performed by modifying the Poisson brackets of the classical theory when the commutators of the quantum theory are modified with the well known relations between both of them and represent a generalization of the treatment shown in \cite{Battisti:2008du,Ali:2014hma,Moumni:2020uki,Atazadeh:2016yeh} by explicitly consider an scalar field with a potential. Let us point out that the modifications found with a perfect fluid describe a bouncing cosmology or an static universe. However as we have shown when the scalar field is taken into account we can have an agreement with the standard inflationary pictures both as an approximation scheme with slow roll parameters and even with analytical solutions. Different approaches to incorporate a GUP have also found agreement with the standard inflationary cosmology before \cite{Cai:2005ra,Zhu:2008cg,Giardino:2020myz,Paliathanasis:2015cza,Giacomini:2020zmv} but not using the classical limit performed in the present article.

We classified the different forms that the GUP can take based on the dependence of each function and present each analysis separately creating cases I to IV treated in sections \ref{S-Case1} to \ref{S-Case4} respectively.

In the first case when the function defining the GUP only depends on the coordinates or in the third case when the dependence is only through the total momentum, a general form of the Friedmann equation was found as well with an equation of motion for the scalar field. With these equations we were able to investigate what is the  modification to the standard scalar field inflation framework when the GUP is present. It was found that we can obtain an exponentially expanding universe if we choose correctly the defining function and we parametrize the approximations needed with three slow roll Hubble parameters instead of the standard two. However the way in which these parameters relate to the standard slow roll parameters is greatly modified opening the window for more forms to realize inflation since in principle we can have inflation even if the standard slow roll parameters are not very small. We showed how these modifications work out in a simple scenario with an exponential potential and found that the $\epsilon$ Hubble parameter is less restrictive in the GUP case than in the standard case.

Furthermore it was shown that for every form of the GUP, whatever its dependence, being diagonal or even non diagonal, there is an analytical solution that can be found. The system of equations treated was always composed of the hamiltonian constraint and the four Hamilton equations of motion modified by the GUP. In all cases a solution describing an expanding universe with constant Hubble parameter can be found by constricting the functions of the GUP to be exponential functions with time. In cases I and III the scalar field is found to be linear with time and thus the potential takes the form of an exponential function with the scalar field. The Hubble parameter is described by two parameters, one coming from the GUP that depend on $\gamma$ and one coming from the potential that can be identified with the cosmological constant. Let us remark that in the standard case there is only one analytical solution describing an exponentially expanding universe, namely the solution when the kinetic term of the field vanishes and the potential is only the cosmological constant. However when the GUP is taken into account we can found a solution that generalizes this setup in the form of an scalar field  with its potential. What is more, even if the cosmological constant is set to vanish when the GUP vanishes, there is still an expanding universe when the GUP is relevant. In this form we can think of the parameter $\gamma$ as varying with the size of the universe in such a way that in the very early universe $\gamma\neq0$ and it is not necessary to incorporate any other matter content, the universe will expand because of the GUP. Then when the universe has a significant size we can neglect the parameter $\gamma$ and thus the scalar field vanishes and it only remains the cosmological constant or only gravity in such a way that the standard cosmology can follow. It was also found that when the parameter coming from the potential was independent of $\gamma$ in the limit of small $\gamma$ the potential obeys the standard slow roll condition, however if there is a dependence on $\gamma$ leading to an static universe we can have inflation that does not obey the slow roll condition but it can fulfil the dS swampland conjecture. It was found that the only restriction to obtain this solution is to restrict the form of the function defining the GUP in such a way that this function can be written as powers of the scale factor or any of the two momenta or it can be written as an exponential in the scalar field. 

In case II the solution described earlier was found to be present in the same form for the scale factor. However since this case is defined with two functions independent to each other, the function in the scalar field commutator was not restricted in any form. Its importance lies in the form of the potential, since that function has to be an exponential in time but the way in which the scalar field depends on time is restricted by the form of such function, and thus in this case we have the freedom to propose that function in any of the ways already studied if we want to obtain such phenomenological features. 

Finally in case IV the situation got more complicated and it was not possible to find a restriction for the form of the GUP in general. It was possible however through a convenient ansatz to find solutions with the same phenomenology as the one described in the other cases. The only difference is that with the solutions found the scalar field is an exponential with time and thus the potential takes the form of two terms with different powers in the scalar field. 

Let us remark that the phenomenology of the GUPs obtained can only be explored in general in some cases. For cases I and III we obtained a minimum measurable value for the scalar field momentum as in the EUP scenario. Furthermore, for case II we have the freedom to choose the function in order to incorporate any of the features of the GUP or the EUP for the scalar field. However for the  scalar field in all the remaining cases and the scale factor in all of the cases we obtained a new form of the GUP in which on the right hand side there appears an expectation value of a power of a coordinate or a momentum. We point out that this uncertainty relation constitutes a new form of the GUP that arrives from our results and we state that this is the correct form that drives inflation for the cases in which it appears. We can not study the phenomenological behaviour in general since we can not relate the expectation value with the uncertainties in general, however we can say a few things about it. We can make the assumption that for some states and for some appropriate values of the constants we can define functions $I(x)$ to relate these expectation values with the corresponding uncertainty, then if $I(x)/x$ has a mininimum we can obtain a minimim measuable value for the coordinate or momentum, depending on the case under consideration. In this way, in all cases we can obtain minimum measurable values for the coordinates and their corresponding momenta when the Hubble and GUP parameters have an appropriate form and the expectation values are computed with appropriate states. Thus, this new form of the GUP can produce the standard phenomenology in appropriated circumstances. The results are summarized in the table below. 

\begin{center}
\begin{tabular}{|c||c|c|c|c|} \hline
	&Functions & Uncertanty Relations  &  Phenomenology \\
	\hline \hline
	Case I &  $F(a,\phi)$&$\Delta a \Delta P_a \geq c_{1} |\langle a^{\ell_{1}} \rangle|$& Scale factor: New \\ 
	&   & $\Delta \phi \Delta P_\phi \geq  (1 + \gamma^2  (\Delta \phi)^2)/2$ & Scalar field: EUP \\
	\hline \hline
	Case II & $F_1(a,P_a)$& $\Delta a \Delta P_a \geq c_{1} |\langle a^{\ell_{1}} \rangle|=c_{2} |\langle P_a^{\ell_{2}} \rangle|$ & Scale factor: New   \\
	&  $F_2(\phi,P_\phi)$  &  $\Delta \phi \Delta P_\phi \geq |\langle F_{2}\rangle|/2$  & Scalar field: GUP or EUP \\
	\hline \hline 
	Case III  & $F(a,\phi,P_a,P_\phi)$ &  $\Delta a \Delta P_a \geq c_{1} |\langle a^{\ell_{2}} \rangle|=c_{2}|\langle P_a^{\ell_{2}} \rangle|$& Scale factor: New \\
	&    & $\Delta \phi \Delta P_\phi  \geq  (1 + \gamma^2  (\Delta \phi)^2) = c_{3} |\langle P_\phi^{\ell_{3}}\rangle |$ & Scalar field: EUP and New\\
	\hline \hline
	Case IV & $F(a,\phi,P_a,P_\phi)$  &  $\Delta a \Delta P_a \geq c_{4} |\langle a^{\ell_{4}} \rangle| = c_{5} |\langle P_a^{\ell_{5}} \rangle|$  & Scale factor: New\\
	& $G(a,\phi,P_a,P_\phi)$  & $\Delta \phi \Delta P_\phi \geq c_{6} |\langle \phi^{\ell_{6}}\rangle| = c_{7}|\langle P_\phi^{\ell_{7}}\rangle |$ & Scalar field: New \\
	&  $F\not= G$  &  $\Delta a \Delta P_\phi \geq d_{1} |\langle a^{\ell_{4}} \rangle| = d_{2} |\langle P_\phi^{\ell_{7}} \rangle|$   & \\
	&    &  $\Delta \phi \Delta P_a \geq d_{3} |\langle \phi^{\ell_{6}} \rangle|= d_4 |\langle P_{a}^{\ell_5} \rangle| $ &\\
	\hline \hline
\end{tabular}
\end{center}

Finally let us remark that all the analytical solutions were found without any approximation and thus it is very interesting that the form in which the standard cosmological constant case is generalized is through the incorporation of an scalar field with a potential and that the only thing that is needed to produce an expansion of the universe is a GUP with an appropriate form. Unfortunately in most of the cases the form of the GUP found was not of the forms studied previously, but they can lead to the behaviours expected with the standard GUP or EUP in appropriate cases. The form of the GUP derived in this article is a new form that have not been studied in the literature but as we have seen, it is the correct form to produce the inflationary period of the universe. Thus we believe it deserves to be explored further to look for phenomenological features that can be derived in general in other contexts. 

 \vspace{1cm}
\centerline{\bf Acknowledgments} \vspace{.5cm} D. Mata-Pacheco would
like to thank CONAHCyT for a grant.

\end{document}